\documentclass[]{article}

\usepackage{geometry}
\usepackage{booktabs}
\usepackage{multirow}
\usepackage{dirtytalk}
\usepackage{caption}
\usepackage{subcaption}
\usepackage{mathtools}
\usepackage{listings}
\usepackage{wrapfig}
\usepackage{comment}

\usepackage[pdftex]{graphicx}
\graphicspath{{figures/}}

\title{Machine Learning Techniques to Detect and Characterise Whistler Radio Waves.}

\author{
	Othniel~J.E.Y.~Konan, Amit~Kumar~Mishra\\
	Department of Electrical and Electronics Engineering\\ University of Cape Town (UCT)\\
	othniel.jean.konan@gmail.com; akmishra@ieee.org
\and
	Stefan~Lotz\\ 
	Science Research and Applications\\
	South African National Space Agency\\
	slotz@sansa.org.za
}
\date{}

\begin{document}

\maketitle

\begin{abstract}
Lightning strokes create powerful electromagnetic pulses that routinely cause very low frequency (VLF) waves to propagate across hemispheres along geomagnetic field lines. VLF antenna receivers can be used to detect these whistler waves generated by these lightning strokes. The particular time/frequency dependence of the received whistler wave enables the estimation of electron density in the plasmasphere region of the magnetosphere. 
Therefore the identification and characterisation of whistlers are important tasks to monitor the plasmasphere in real time and to build large databases of events to be used for statistical studies. The current state of the art in detecting whistler is the Automatic Whistler Detection (AWD) method developed by Lichtenberger (2009) \cite{Lichtenberger2009}. This method is based on image correlation in 2 dimensions and requires significant computing hardware situated at the VLF receiver antennas (e.g. in Antarctica).
The aim of this work is to develop a machine learning based model capable of automatically detecting whistlers in the data provided by the VLF receivers. The approach is to use a combination of image classification and localisation on the spectrogram data generated by the VLF receivers to identify and localise each whistler. The data at hand has around 2300 events identified by AWD at SANAE and Marion and will be used as training, validation, and testing data.
Three detector designs have been proposed. The first one using a similar method to AWD, the second using image classification on regions of interest extracted from a spectrogram, and the last one using YOLO, the current state of the art in object detection. It has been shown that these detectors can achieve a misdetection and false alarm of less than 15\% on Marion's dataset.
\end{abstract}

\section{Introduction}
\begin{figure}[!h] 
	\centering
	\captionsetup{justification=centering}
	\includegraphics[width=.8\textwidth]{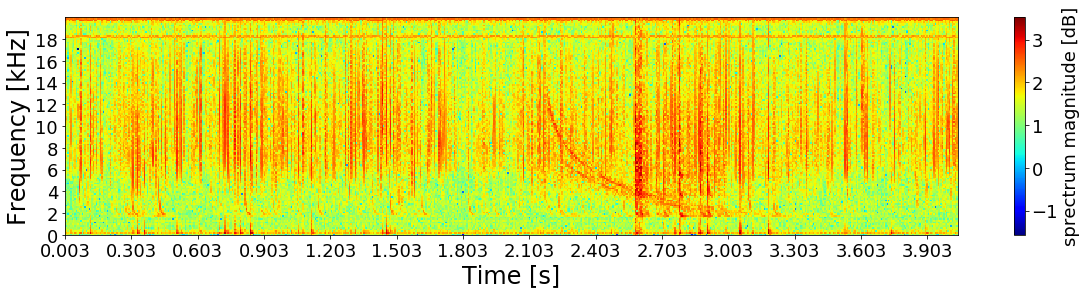}
	\caption{Example of whistler wave trace in spectrogram.}
	\label{fig:ch1_spectrogram_data}
\end{figure}

Lightning strokes create powerful electromagnetic pulses that result in Very Low Frequency (VLF) waves propagating along the magnetic field lines of the earth. Due to the dipole shape of the geomagnetic field, these waves travel upward from the stroke location out through portions of the plasmasphere and back to the Earth's surface at the field line foot point in the opposite hemisphere.  VLF antenna receivers set up at various high and middle latitude locations can detect whistler waves generated by these lightning strokes.
The propagation time delay of these waves is dependent on the plasma density along the propagation path. This enables the use of whistler wave observations for characterising the plasmasphere in terms of particle number and energy density. The dynamics of energetic particle populations in the plasmasphere are an important factor in characterising the risk to spacecraft in orbit around Earth. Annual global lightning flash rates are on the order of 45 flash/s ~\cite{Christian2003}. The resulting high occurrence rate of whistler events makes it impossible to identify and characterise them in a reasonable time.  Therefore the automatic detection and characterisation of whistlers are valuable to the study of energetic particle dynamics in the plasmasphere and to develop models for operational use.\\
Lichtenberger ~\cite{Lichtenberger2009} developed an automatic detector and analyser based on the Appleton-Hartree dispersion relation and experimental models of particle density distribution.  Recent advances in artificial neural network-based image processing methods – for example, convolutional networks~\cite{Goodfellow2016} may be able to provide an alternative method for the automatic identification and characterisation of whistler events in broadband VLF spectra. \\
Model development is based on training a neural-network-based model on a large set of spectrograms
with whistler events identified by the nodes of the Automatic Whistler Detection and Analysis Network (AWDAnet ~\cite{Lichtenberger2008}).  Spectrograms will be presented in the form of images (Figure \ref{fig:ch1_spectrogram_data}) to take advantage of the wide range of image-processing techniques available for this type of object identification.

\subsection{Objectives of Study}
This project investigates the feasibility of finding and implementing methods - including the one proposed by Lichtenberger ~\cite{Lichtenberger2009} - to localise whistler radio waves observed in the plasmasphere. This problem encompasses the following questions:
\begin{enumerate}
	\item[(1)] Is it possible to develop a new method in detecting whistlers different from the SOTA?
	\item[(2)] Can any of the new methods localise the start and end time of the whistler?
	\item[(3)] Do the new methods perform better than the current SOTA?
\end{enumerate}
The purpose of this study is thus to develop and implement a few methods to localise any whistlers in given VLF spectra. By localisation, it is implied that the starting time and ending time of each whistler should be specified within a range of uncertainty.

The proposed method should localise any whistlers in given spectra within 60 seconds of reading the raw data. 

A major limitation of this project is the availability and diversity of the data. The dataset used for this project has 2571 observations in total collected in 2013 with 2196 samples from Marion Island and 375 from SANAE IV. 
Another limitation is the verification of the result of the proposed method which is limited by the labelled data provided by the SOTA. The SOTA provides a time location of each whistler at $5 kHz$ and therefore not the starting point. Thus, it is difficult to use the SO-TA as a benchmark for the evaluation of the starting and ending time of the detected whistler.

\section{Project Requirements}
The new method for whistlers identification is intended to detect and localise whistlers within any given raw VLF data faster than real-time. In our context, detection means finding the number whistlers in the raw data, while localisation means providing a bounding box coordinate of the location of each whistler in a frequency-time plot of the raw data. Following the criteria of AWD, the detection and localisation should be done faster than the collection time of the data on an average computer.    
\subsection{Requirement 1: Whistler Detection and Localisation}
$R_1:$ \textit{The system should be able to count the number and time occurrence of whistlers in a data file.}\\
The particular time-frequency dependence of the received whistler wave enables the estimation of electron density in the plasmasphere region of the Earth's magnetic field. The data files provided by the South African National Space Agency (SANSA) contain at least one target over per  observation. The whistlers can be visually seen on the data file spectrogram as seen as curved lines in the spectrogram in Figure \ref{fig:ch1_spectrogram_data}. So far, the AWD algorithm is capable of detecting the whistlers in a data file. The new method will be developed based on a set of whistlers identified by the Automatic Whistler Detection (AWD) algorithm developed by Lichtenberger (2009). The AWD method is a physics-based method that is the current state of the art in automatic detection of whistlers \cite{Lichtenberger2008}. \\
The system requirement for requirement 1 (R1) are tabulated in Table \ref{tab:ch3_system_req1}. 
R1 entails that a software must be written that given a data file, outputs the location (starting time and duration) of whistlers as well as a figure of merit on the accuracy of the detection.
\begin{table}[!h]
	\centering
	\caption{System requirements for requirement 1 ($R_{1x}$ where x is a sub-requirement)}
	\label{tab:ch3_system_req1}
	\begin{tabular}{@{}cl@{}}
		\toprule
		\textbf{Requirements} & \textbf{Description} \\ 
		\midrule
		$R_{11}$ & \begin{tabular}[c]{@{}l@{}}Create software that given a data file (.vr2 file) \\ ouputs a list of detected whistlers\end{tabular} \\ 
		\midrule
		$R_{12}$ & \begin{tabular}[c]{@{}l@{}}Each whistler in the output must\\  be followed by a start and end time\end{tabular} \\ 
		\midrule
		$R_{13}$ & \begin{tabular}[c]{@{}l@{}}Each whistler event in the output file\\  must have a figure of merit for its accuracy\end{tabular} \\ 
		\bottomrule
	\end{tabular}
\end{table}

Since the system will be trained from the detection made by the AWD algorithm, the accuracy of the detection of the system will be tested by comparing its performance to the one of the AWD on a large set of data.

\subsection{Requirement 2: Faster than real time}
$R_2:$ \textit{The system should process any data faster than the generating time of the data, \textit{i.e.} a data of length $t$ should have a processing time of $t_{processing} \leq t$.}\\

Global annual average lighting strokes are about 45/second. Ideally, the system should be able to process 45 data files per second. The focus of this project is to evaluate the performance of machine learning compared to the existing AWD method, therefore, real-time processing is not a point of emphasis. It is assumed that a report on the whistlers occurrence is required at the end of every day, therefore, a 24 hours delay is assumed to be the real-time of the system.\\
The system requirement for requirement 2 (R2) are tabulated in Table \ref{tab:ch3_system_req2}. The system has 24 hours to process the data collected in a day, therefore, it is assumed that the average processing time for a sample of duration $t$ should be less than $t$. 
\begin{table}[!h]
	\centering
	\caption{System requirements for requirement 2}
	\label{tab:ch3_system_req2}
	\begin{tabular}{@{}cl@{}}	\toprule
		\textbf{Requirements} & \textbf{Description} \\ 
		\midrule
		$R_{2}$ & \begin{tabular}[c]{@{}l@{}}Data processing time should be \\ less than collection time\end{tabular} \\
		\bottomrule 
	\end{tabular}
\end{table}

The system is to be implemented as a computer program. The CPU time will be used to determine the processing time $t_{processing}$ of the detection. The acceptance strategy consists of evaluation the processing ration defined as:
$$r=\frac{t_{processing}}{t}$$
with $t$, the duration of the sample. A ratio less or equal to one implies that the processing time is smaller than the collection time of the sample.\\
The specifications of the whistler detector and localiser are defined in Table \ref{tab:ch3_system_specs}. This research focuses on the ability of new methods to detect whistler radio waves. Finner detection resolutions are not of interest since the focus is on the presence of the radio waves. Therefore, the resolution of the detector should be 100 milliseconds per detection. The AWD achieved a misdetection of 10\% and a false alarm of between 20 to 50\% \cite{Lichtenberger2008}, the goal in this research is to achieve less than 20\% for both misdetection and false alarm rate. Finally, the detection should happen in real-time, which implies that the processing time should be less than the collection time.
\begin{table}[ht]
	\centering
	\caption{System specifications}
	\label{tab:ch3_system_specs}
	\begin{tabular}{@{}cl@{}}
		\toprule
		\textbf{Specification} & \textbf{Description} \\ \midrule
		$S_{1}$ & \begin{tabular}[c]{@{}l@{}}The resolution of the detection should be \\ 100ms per detection\end{tabular} \\ \midrule
		$S_{2}$ & \begin{tabular}[c]{@{}l@{}}The false alarm and misdetection rate of the \\ whistlers should be less than 20\%.\end{tabular} \\ 
		\midrule
		$S_{3}$ & \begin{tabular}[c]{@{}l@{}}The detection of whistlers from a data file \\ should be done faster than the time duration \\ of the data\end{tabular}\\
		\bottomrule
	\end{tabular}
\end{table}

\section{Whistler Data}
\begin{figure}[!h]
	\centering
	\begin{subfigure}[c]{\textwidth}
		\centering
		\includegraphics[width=.8\linewidth]{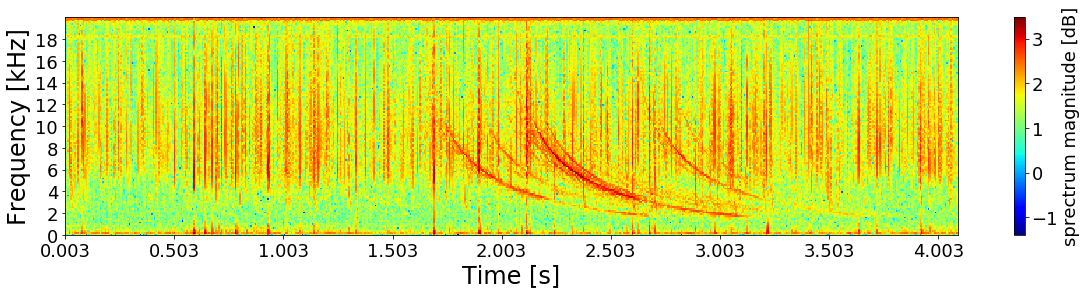}
		\caption{Data sample from Marion sampled at 40kHz}
		\label{fig:ch4_data_partition_marion}
	\end{subfigure}
	\newline
	\begin{subfigure}[c]{\textwidth}
		\centering
		\includegraphics[width=.8\linewidth]{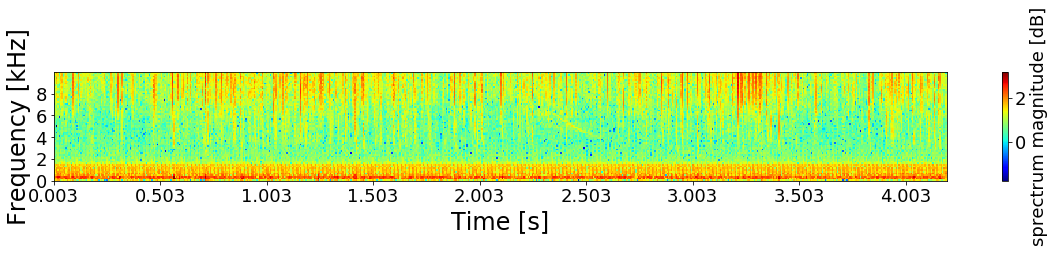}
		\caption{Data sample from Marion sampled at 22kHz}
		\label{fig:ch4_data_partition_sanae}
	\end{subfigure}
	\caption{Samples from data collected at Marion and SANAE IV}
	\label{fig:ch4_data_partition}
\end{figure}

The development of the whistler detector method presented by Lichtenberger \textit{et al.} \cite{Lichtenberger2008} led to the creation of a global network of Automatic Whistlers Detector and Analyser (AWDA) known as Automatic Whistlers Detector and Analyser Network (AWDAnet) which is made of AWD nodes monitoring the plasmaspheric electron density variation in real-time. The data provided for this research was collected at two nodes, one at Marion island in 2013 (2196 files with data sampled at 40kHz) and another at SANAE IV in 2012 (375 files with data sampled at 20kHz). Each data file contains the value of the signal received by the North/South (NS) and East/West (EW) pointing orthogonal loop antennas used at the site.\\
Figure \ref{fig:ch4_data_partition_marion} and Figure \ref{fig:ch4_data_partition_sanae} present respectively, the spectrogram of one sample from Mario and SANAE IV.
The data collected at Marion has frequency only up to 20 kHz while the one collected at SANAE IV has a maximum frequency of 10kHz. Unfortunately, the data collected at SANAE IV suffered from low SNR caused by a problem with the AWD amplifier. On the assumption that whistlers have the same characteristics and only the noise varies from one site to another, the SANAE IV data will not be used for the preliminary design of the whistler detector but for testing.  \\
For the design of the whistler detector, the data at hand is split into two main portions, a training set and a testing set with a portion of $67\%$ and $33\%$ in other words $1471$ training samples and $375$ samples.\\
\subsection{Whistler Simulation}
\begin{figure}[!h]
	\centering
	\includegraphics[width=0.5\textwidth]{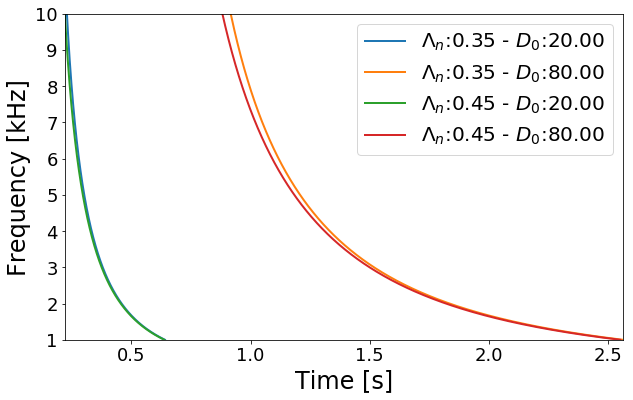}
	\captionsetup{justification=centering}
	\caption{Whistlers' dispersion approximation with a nose frequency of $f_n=25kHz$, and at varying zero dispersion and normalised electron gyrofrequency.}
	\label{fig:ch4_wshitler_approx}
\end{figure}

L. C. Bernard \textit{et al.} \cite{Bernard1973} proposed a equation for approximating the whistler's dispersion. From this, an expression is derived relating the nose frequency $f_n$, the zero dispersion $D_0$, the normalised electron gyrofrequency $\lambda_n$, the time travel $t$ and the frequency $f$ of the whistlers:
\begin{equation}
t = \frac{D_0}{(1+\Lambda_n)\sqrt{f}}\frac{(1+\Lambda_n)-(3\Lambda_n-1)(f/f_n)}{1-\Lambda_n (f/f_n)}
\label{eq:whistler_dispersion_approximation}
\end{equation}
To simulate the whistlers, we make use of this equation to generate a time-frequency representation of the whistler at the chosen parameters. \\
Equation \ref{eq:whistler_dispersion_approximation} has four parameters:
\begin{itemize}
	\item \textbf{$f$}: The frequency range of the whistler. Since most of the peak of the whistler are located between 3 to 6kHz, a frequency range is chosen such that $1kHz \leq f \leq 10 kHz$.
	\item \textbf{$f_n$}: The nose frequency of the whistler. Lichtenberger \textit{et al.} \cite{Lichtenberger2008} selected a nose frequency of $f_n=25kHz$, since our work is to be compared to theirs, this nose frequency is chosen.
	\item \textbf{$D_0$}: The zero dispersion of the whistler is chosen to match the one used by Lichtenberger \textit{et al.} \cite{Lichtenberger2008}. The dispersion is thus chosen such that $20s^{1/2} \leq D_0 \leq 80s^{1/2}$ 
	\item \textbf{$\Lambda_n$}: The normalised electron gyrofrequency is chosen such that $0.35 \leq \Lambda_n \leq 0.45$ as per Figure I in Bernard 1973 \cite{Bernard1973}.
\end{itemize}
Figure \ref{fig:ch4_wshitler_approx} presents the whistler approximation at the boundaries of the variable parameters of Equation \ref{eq:whistler_dispersion_approximation}.
For every frequency point, the time travel increases linearly as a function of the zero-dispersion with an increase in the zero-dispersion increasing in the time travel. The fastest dispersions occur at $D_0=20s^{1/2}$ and the slower ones at $D_0=80s^{1/2}$. The normalised electron gyrofrequency $\Lambda_n$ slightly affects the time travel with an increase in $\Lambda_n$ resulting in a decrease in the whistler's time travel.\\

\begin{figure}[!h]
	\centering
	\includegraphics[width=0.4\textwidth]{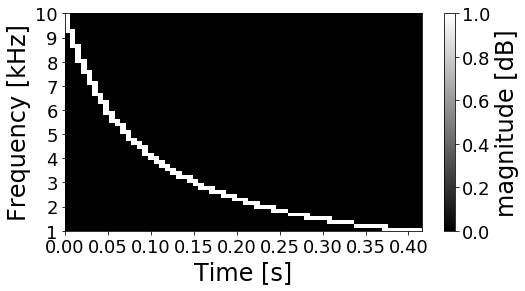}
	\caption{Simulated whistler with $fn=25kHz$, $D_0=20s^{1/2}$, $\Lambda_n=0.35$, and $1kHz \leq f \leq 10kHz$}
	\label{fig:ch4_whistler_sim}
\end{figure}

The approximations in Figure \ref{fig:ch4_wshitler_approx} are in two dimensions and must be converted to three-dimensional data so that they can be visualised as spectrograms. To achieve this, a three-dimensional array is created where the first and second axis represents the time and frequency obtained from the approximation. Since the approximation is of the form, time as a function of frequency, the time and frequency have the same number of sample. To fix this problem, the time and frequency are scaled based on the time and frequency resolution of the spectrogram of the whistlers data. The third dimension represents the magnitude of the whistler which follow the dispersion based on the scaled time and frequency. Figure \ref{fig:ch4_whistler_sim} shows the spectrogram of the simulated whistler with $fn=25kHz$, $D_0=20s^{1/2}$, $\Lambda_n=0.35$, and $1kHz \leq f \leq 10kHz$.

\subsection{Data Preprocessing}
The primary purpose of preprocessing the data is to reduce the noise (whistler, VLF transmissions, transient signal, etc \ldots) in the spectrogram without removing any whistler information. Ideally, if the noise is removed, finding the shape of a signal (whistlers) in another (spectrogram) could be solved using simple methods such as matched filtering. 
\subsubsection{Z score transform}
In the domain of statistics, the Z score or Standard score of a sample is the signed number quantifying how many standards deviation the sample is from the mean of the observed series \cite{kreyszig11}. If both the mean and standard deviation of a series are known, the z score is expressed as:
\begin{equation}
z_i = \frac{x_i-\mu}{\sigma}  
\label{eq:zscore}
\end{equation}
where
\begin{description}
	\item $z_i$ and $x_i$ respectively the z-score and the value of data point $i$ \item $\mu$ and $\sigma$, the mean and standard deviation of the data
\end{description}
and the resulting series has two interesting properties:
\begin{itemize}
	\item[(1)]  a mean of 0
	\item[(2)]  a standard deviation of 1
\end{itemize} 

\subsubsection{Detrending}
Detrending is the process of removing any known trend in a series. Similarly to the Z score transform, detrending can be used to reduce the presence of spherics and transient by either removing the mean or the linear least square of each frequency and time cut.\\
\textbf{Reduction by the Mean}\\
An approach to detrending is to remove the mean of the series from each element. Assuming that $x$ is an element of the series $X$, the transform $D_c{}$ which detrends a series by a constant c is given by Equation \ref{eq:detrending_tranform}.
\begin{equation}
x_c = D_c\{x\} = x-c 
\label{eq:detrending_tranform}
\end{equation} 
The series $X_c$ such that $X_c = D_c\{X\}$ has a mean of $\mu_c=\mu-c$ and a standard deviation of $\sigma_c = \sigma$. 
Detrending by the mean thus result in a series $X_\mu$ such that $X_\mu = D_\mu{X}$ with a mean of $\mu_\mu=0$ and a standard deviation of $\sigma_\mu = \sigma$.\\
\textbf{Linear Least Square}\\
Reducing each cut by its linear least-square fit as opposed to its mean does not necessarily centre the series at zero. This would be the case if the linear least-square fit has a gradient of zero (constant through all increment). That could likely occur if the series contains spherics or transients. However, if the fit has a gradient, which is the case for series containing whistlers, the series is not centred to zero. It thus becomes inconvenient or difficult to find a simple relation between the result of the transform and the original spectrogram.

\subsection{Evaluation of the techniques}
\begin{figure}[!h] 
	\centering
	\includegraphics[width=.5\textwidth]{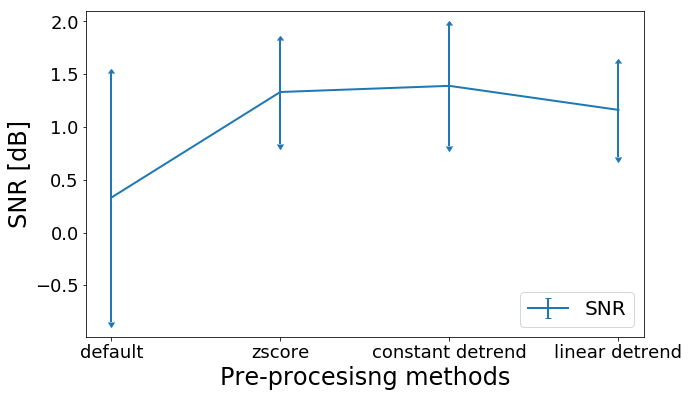}
	\caption{SNR of each pre-processing method on the training set. }
	\label{fig:ch4_preprocess_snr}
\end{figure}

To evaluate the performance of the preprocessing methods, all samples in the training set were preprocessed and the overall SNR per method was calculated \footnote{Since  these techniques can scale the spectrogram's magnitude, each spectrogram is scaled using $\mu \pm 4\sigma$ dB - with $mean$ and $\sigma$ obtained from the distribution of the spectrogram from Marion - before SNR calculation.}. The results of these methods on the training set are shown in Figure \ref{fig:ch4_preprocess_snr}. 
The overall SNR of the training set after applying any of the pre-processing transforms is higher than the default SNR. Among those, the constant detrending has the highest SNR, just 0.06dB above the second-highest, the Z score. However, the standard deviation of the Z score is 0.08db below the one of the constant detrending. The Z score will thus be the default transform in the comming sections.

\section{Detection Using Cross-Correlation with a Simulated Whistler}
\begin{figure}[!h] 
	\centering
	\includegraphics[width=.8\textwidth]{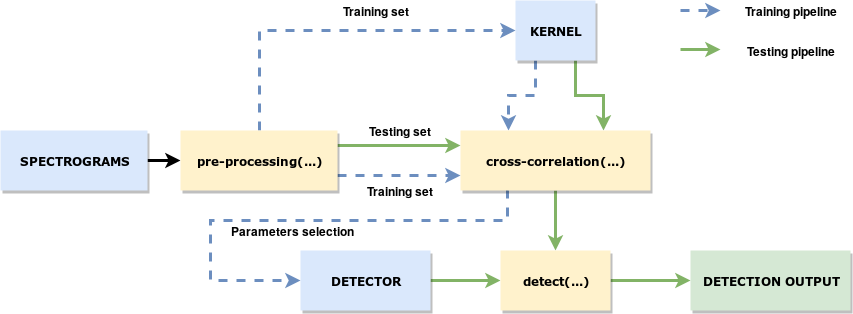}
	\captionsetup{justification=centering}
	\caption{Overview of the whistler detection pipeline using cross correlation with a whistler kernel.}
	\label{fig:ch5_design}
\end{figure}

The best and so far only approach in the whistler detection problem is from Lichtenberger \textit{et al.} \cite{Lichtenberger2008}.In his paper, Lichtenberger proposes a method which makes use of a simulated whistler using the Bernard approximation of whistlers \cite{Bernard1973} and a square-law detector. Our first approach in detecting whistler waves is to create a similar method to the one proposed by Lichtenberger \textit{et al.}\\
The detector's design can be broken down in two pipelines as presented in Figure \ref{fig:ch5_design}.

The method presented in Figure \ref{fig:ch5_design} has three processes and three inputs. The first process (\textit{preprocessing(...)}) is preprocess the spectrogram to increase its SNR. The second process (\textit{corr-correlation(...)}) consists of finding the whistler's shape in the spectrogram of interest. The third process (\textit{detect(...)}), consists of applying a suitable threshold to differentiate the whistlers from the noise present in the spectrogram. These processes are used in the training and testing pipeline of the design. During training, samples from the training set are preprocessed and analysed to select the kernel which later is used in conjunction with the same training samples to design the detector. During testing, the kernel is cross-correlated with pre-processed samples and the designed detector is applied to this result to generate the output of the detection.

\subsection{Kernel Selection}
\begin{figure}[!h] 
	\centering
	\includegraphics[width=.8\textwidth]{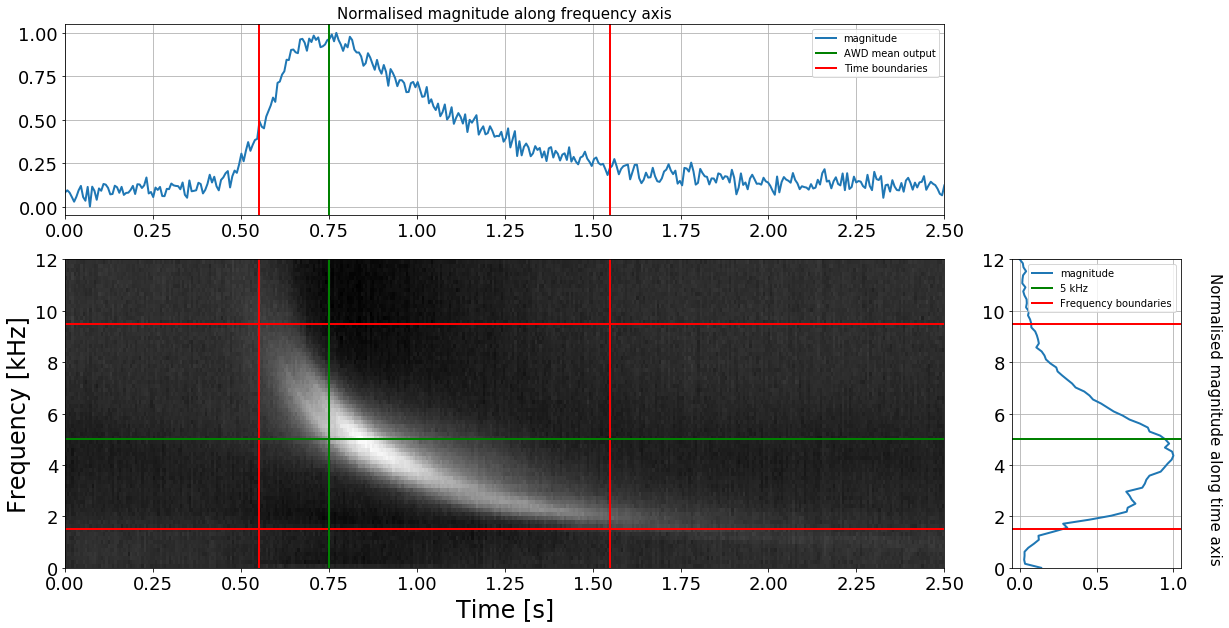}
	\captionsetup{justification=centering}
	\caption{Mean of whistler cuts alongside the sum along the time and frequency axis. The boundaries chosen for the whistler are 0.55s, 1.55s, 1.5kHz and 9.5kHz}
	\label{fig:ch5_kernel_time_freq}
\end{figure}

Each sample in the dataset contains at least one whistler. The first approach is to use whistlers whose characteristics are intrinsic to the data at hand. Thus the labels provided by the AWD are used to extract whistlers from each sample of the training set. This extraction consists of generating whistler cuts from each spectrogram such that the entirety of the visible section of each whistler is present in its corresponding cut. We, therefore, choose these cuts to be 2.5 seconds long with the AWD output (the 5kHz time) at 30\% of 2.5s and ranging from 0 to 12 kHz. Moreover, since only the whistlers are of interest, each sample is pre-processed using the Zscore before generating the cuts.\\
The mean of these cuts is observed Figure \ref{fig:ch5_kernel_time_freq}. One main whistler can be observed in this cut. While this shows the intrinsic nature of the whistlers present in the training set, the boundaries initially chosen for the cuts contains portions of the spectrogram that are not of interest. We thus observed the magnitude of these cuts along the time and frequency axis. 
The plot of the normalised magnitude along the frequency axis shows the AWD output at 0.75s and the selected time boundaries at 0.55s and 1.55s. The normalised magnitude along the time axis shows the 5kHz reference line and the two selected boundaries at 1.5kHz and 9.5kHz. The final whislter is thus the portion of this cut delimited by these boundaries.\\

The whistler extracted from the data can be used to simulate a whislter whose characteristics fit the data. The Bernard approximation of the Whistler dispersion provides two key parameters, the normalised nose frequency and the zero dispersion, both characterising the shape and duration of the whistlers. From Equation \ref{eq:whistler_dispersion_approximation}, $t$  can be written as a linear function of $$t = f(f_n,f,\Lambda_n)D_0$$ with $f_n$, $\Lambda_n$, and $f$ fixed parameters. Since $20 \leq D_0 \leq 80$, an increase in $D_0$ results in an increase in the duration of the approximated whistler. 

\begin{figure}[!h] 
	\centering
	\includegraphics[width=.5\textwidth]{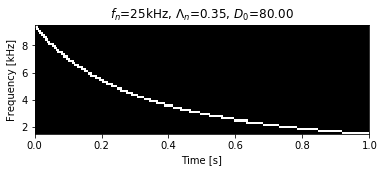}
	\captionsetup{justification=centering}
	\caption{Kernel generated with $f_n=25kHz$, $\Lambda_n=0.35$, $D_0=80$, $1.5kHz \leq f \leq 9.5kHz$ with duration of respectively $t=1.03$ seconds. }
	\label{fig:ch5_kernel_bernard}
\end{figure}

Since we have control over the shape and mostly, the duration of the whistlers, few kernels spanning over the range of D0 could be used. For example, Figure \ref{fig:ch5_kernel_time_freq} shows the presence of a whistler lasting for more or less one second, this implies that the whistlers present in the data have more or less duration of one second. As a result, a simulated whistler with a duration of around one second should be optimal. The simulated whistler corresponding to these characteristics is shown in Figure  \ref{fig:ch5_kernel_bernard}. 

\subsection{Spectrogram preprocessing}
\begin{figure}[!h] 
	\centering
	\includegraphics[width=\textwidth]{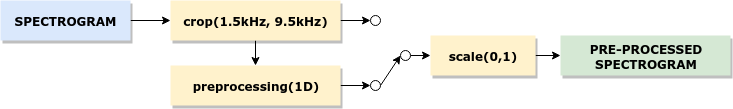}
	\captionsetup{justification=centering}
	\caption{Pre-processing pipeling. Each spectrogram is firstly cropped, can be preprocessed using one of the 1D techniques, and lastly is scaled between 0 and 1.}
	\label{fig:ch5_preprocessing}
\end{figure}

The whistlers in spectrograms can be emphasised by applying different preprocessing techniques. As a result, each spectrogram undergoes preprocessing before being correlated with the kernel of choice. Moreover, the correlation process aims to generate a one dimensional output, therefore, each spectrogram must have the same frequency range as each kernel. We thus crop each spectrogram from 1.5kHz to 9.5kHz as shown in the preprocessing pipeline illustrated in Figure \ref{fig:ch5_preprocessing}.  
After cropping, the spectrogram can furthermore be processed using either a one or two-dimensional pre-processing methods. The last stage of preprocessing consists of scaling the spectrogram to withing the range of 0 and 1.

\subsection{Cross-Correlation}
\begin{figure}[!h]
	\centering
	\begin{subfigure}[c]{.8\textwidth}
		\centering
		\includegraphics[width=\textwidth]{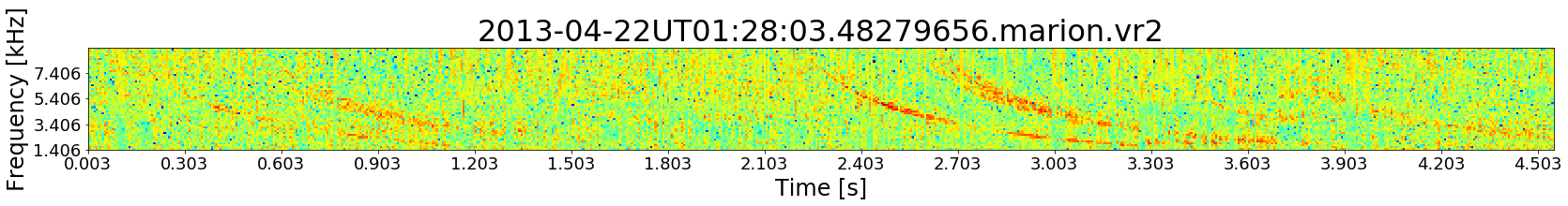}
		\captionsetup{justification=centering}
		\caption{Pre-processed spectrogram with whistlers starting at $t=[0.13,0.55,2.2,2.64,2.8,3.28]s$.}
		\label{fig:ch5_spec_cropped}
	\end{subfigure}
	\begin{subfigure}[c]{.8\textwidth}
		\centering
		\includegraphics[width=\textwidth]{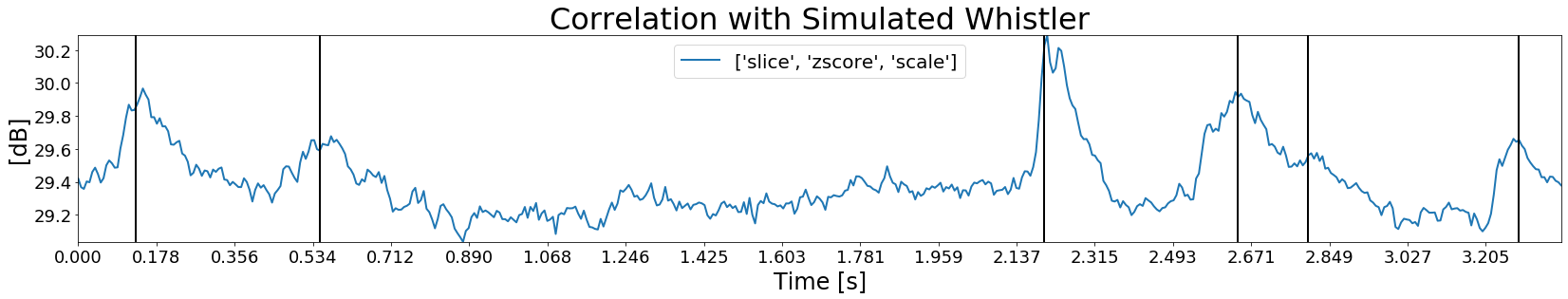}
		\captionsetup{justification=centering}
		\caption{Correlation using the simulated kernel (Figure \ref{fig:ch5_kernel_bernard}) showing local maxima at $t=[0.13,0.55,2.2,2.64,2.8,3.28]s$.}
		\label{fig:ch5_corr_kernel_sim}
	\end{subfigure}
	\captionsetup{justification=centering}
	\caption{Results of correlation using the simulated whislter}
	\label{fig:ch5_corr_kernels}
\end{figure}

The second process of the pipeline in Figure \ref{fig:ch5_design} is the cross-correlation between the preprocessed spectrogram and the kernel of choice. This process measures the similarities between the spectrogram and the kernel over time. Since the integrity of the results matter, only a valid cross-correlation is used. Moreover, it is easier to work with a 1D matrix, thus the kernel and the preprocessed spectrogram must have the same number of frequency sample.\\
Figure \ref{fig:ch5_corr_kernel_sim} displays the result of this process on the preprocessed spectrogram in Figures \ref{fig:ch5_spec_cropped} with the kernels presented in Figure \ref{fig:ch5_kernel_bernard}.

Figure \ref{fig:ch5_corr_kernel_sim} has five local maxima at $t=[0.13,0.55,2.2,2.64,2.8]s$ and only captures the first 3.38 seconds of the original spectrogram. The result has local maxima at the location of the first four whistlers present in Figure \ref{fig:ch5_spec_cropped}. 

\subsection{Static Detector}
\begin{figure}[!h]
	\centering
	\begin{subfigure}[t]{.35\textwidth}
		\includegraphics[width=\textwidth]{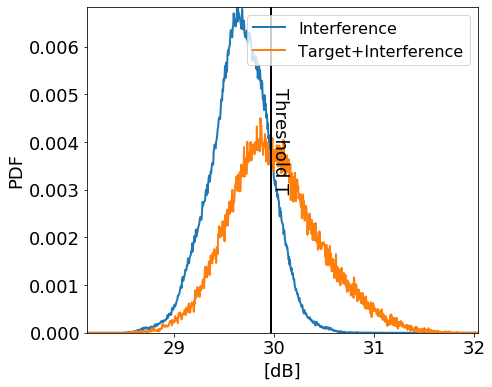}
		\captionsetup{justification=centering}
		\caption{Probability Density Function ($PDF$) of the of the interference and target plus interference.}
		\label{fig:ch5_detector_pdf}
	\end{subfigure}
	\begin{subfigure}[t]{.35\textwidth}
		\includegraphics[width=\textwidth]{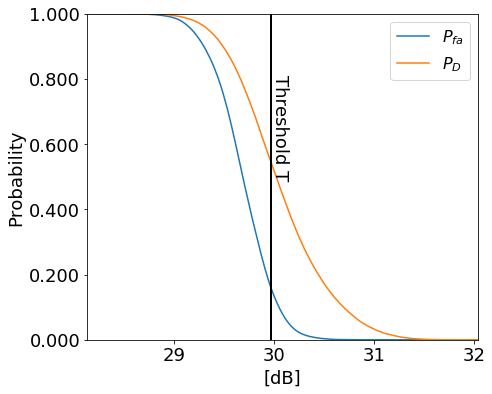}
		\captionsetup{justification=centering}
		\caption{Probability of False Alarm $P_{fa}$ and Probability of Detection $P_D$ per threshold.}
		\label{fig:ch5_detector_pfa_pd}
	\end{subfigure}
	\captionsetup{justification=centering}
	\caption{Statistics of interference and target plus interference of the cross-correlation on the training set with threshold $T=29.97dB$.}
	\label{fig:ch5_detector_stats}
\end{figure}

The third input to the pipeline in Figure \ref{fig:ch5_design} is the detector. The role of the detector is to generate a suitable threshold to discriminate targets from interference. To understand the behaviour of the targets and interference in the system, the interference ($I$) and target plus interference ($T+I$) from all cross-correlations resulted using the training set are extracted.\\
Figure \ref{fig:ch5_detector_stats} presents the probability distribution of both interference and target plus interference as well as the probability of false alarm ($P_{fa}$) and probability of detection ($P_D$) per threshold.\footnote{$P_{fa}$ is the area under the interference $PDF$ curve to the right of the threshold while $P_D$ is the area under the target+interference $PDF$ curve to the right of the threshold }

Considering all training samples, the interference has a mean of $\mu_I=29.68dB$ and a standard deviation of $\sigma_I=0.30dB$ and the targets plus interference has a mean of $\mu_{T+I}=30.06dB$ and a standard deviation of $\sigma_{T+I}=0.47dB$. The $T+I$ and $I$ PDF curves are both Rayleigh distributed with the $TI$ PDF curve flatter and skewer to the right. Selecting a static threshold for target detection can thus be easily done by selecting a desired $P_d$ or $P_{fa}$. A simple approach is to select $T$ where $P_D-P_{fa}$ is the highest. In this case, it corresponds to $T=29.97dB$ where $P_D=5.44\mathrm{E}{-1}$ and $P_{fa}=1.58\mathrm{E}{-1}$ as shown in Figure \ref{fig:ch5_detector_stats}.

\begin{figure}[!h]
	\centering
	\begin{subfigure}[c]{.8\textwidth}
		\centering
		\includegraphics[width=\textwidth]{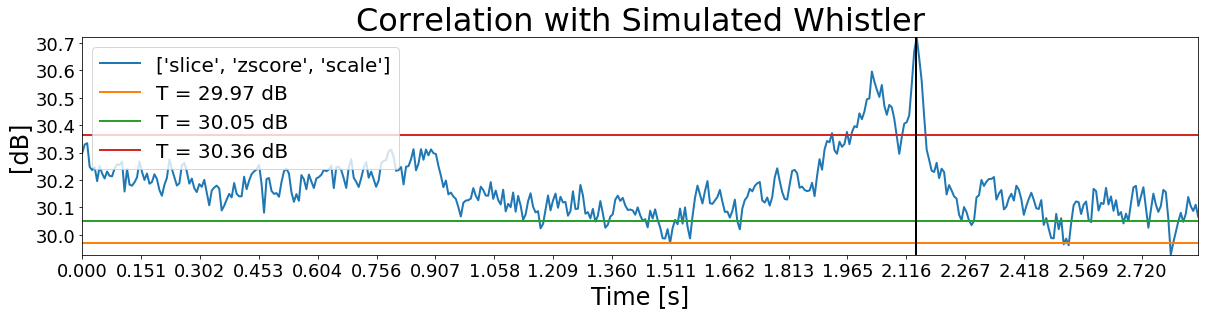}
		\captionsetup{justification=centering}
		\caption{Cross-correlation with whistler present at 2.14s.}
		\label{fig:ch5_detector_static_corr1}
	\end{subfigure}
	\begin{subfigure}[c]{.8\textwidth}
		\centering
		\includegraphics[width=\textwidth]{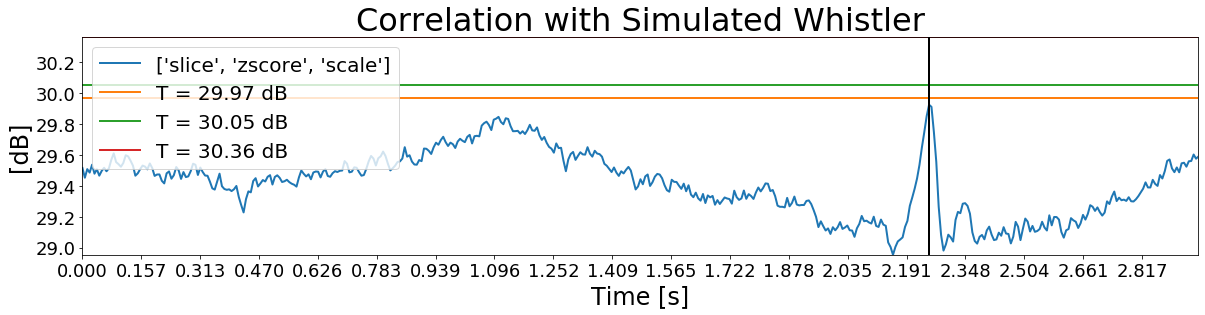}
		\captionsetup{justification=centering}
		\caption{Cross-correlation with whistler present at 2.25s.}
		\label{fig:ch5_detector_static_corr2}
	\end{subfigure}
	\captionsetup{justification=centering}
	\caption[Cross-correlation result with static thresholds]{Cross-correlation result with static thresholds selected such that $(T,P_{fa}) \in \{(29.9,71.58\mathrm{E}{-1}),(30.05,1\mathrm{E}{-1}),(30.36,1\mathrm{E}{-2})\}$.}
	\label{fig:ch5_detector_static}
\end{figure}

Figure \ref{fig:ch5_detector_static} shows a few selected thresholds alongside the result of the cross-correlation for two training samples.

In a system with homogeneous interference, a static threshold can be used to detect targets from interference. However, the interference after pre-processing of the spectrograms and cross-correlation are unpredictable. For example, the interference presents in Figure \ref{fig:ch5_detector_static_corr2} changes magnitude at different section of the cross-correlation. Moreover, the whistler's maximum cross-correlation value is below the threshold level guarantying the maximum difference between $P_D$ and $P_{fa}$. In Figure \ref{fig:ch5_detector_static_corr1}, however, the interference, as well as the target plus interference, are well above this threshold. This problem cannot be efficiently tackled if using a static threshold detector. We thus propose the use of adaptive threshold-based detectors such as Constant False Alarm Rate  (CFAR) detectors.\\

\subsection{CFAR Detectors}
\begin{figure}[!h] 
	\centering
	\includegraphics[width=.8\textwidth]{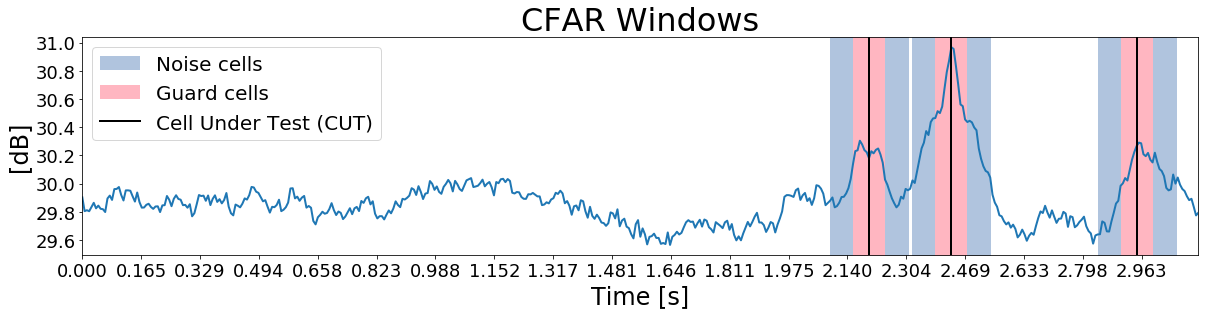}
	\captionsetup{justification=centering}
	\caption{CFAR windows.}
	\label{fig:ch5_detector_cfar_window}
\end{figure}

CFAR detectors require at least three parameters, the window size which includes the cell under test (CUT), the noise cells and guard cells, and a scaling factor given as a function of the probability of false alarm.\\
CFAR detectors use a window and a statistical scaling factor to generate an adaptive threshold. One challenge with CFAR is the selection of the CFAR window size. The window size $W$ is dependent of the number of guard cells $G$ and the number of noise cells $N$ on each side of the cell under test, precisely, $W = 2(N+G)+1$. Ideally, the width of the local maxima should be less than twice the number of guard cell ($whistler\leq2G$) while the next consecutive whistler should be located at $\pm (N+G)$ cells away from the CUT. However, whistlers do not have a constant or minimum duration from one to the next. Moreover, the presence of whistlers might be faint in the spectrograms resulting in some cases where the interference levels are above the whistlers. \\
Figure \ref{fig:ch5_detector_cfar_window} shows the window selected by the CFAR detector. With each cell being 6.4ms long, the window selected in Figure \ref{fig:ch5_detector_cfar_window} has 7 guard cells (44.77ms) around each CUT followed by 10 noise cells (63.95ms) accounting for a total of 35 cells (223.83ms) in each window.  

The most basic type of CFAR is the Cell Averaging CFAR where the adaptive threshold at the CUT ($T_{CUT}$) is given by 
$$[T_{CUT}]_{dB} - [\hat{\sigma}_{CUT}^2]_{dB} = [T]_{dB}$$
Let $X_{dB}$ be the difference between the threshold and the noise power, then
\begin{equation}
P_{fa} = \left(\cfrac{1}{1+\cfrac{10^{X_{dB}/_{10}}}{2N}}\right)^{2N}
\label{eq:cfar_pfa_XdB}
\end{equation}
We previously mentioned how $P_{fa}$ can be obtained by selecting a threshold using Figure \ref{fig:ch5_detector_pfa_pd}, however, since a single whistler in the cross-correlation can cover many target cells, this method is thus difficult to visually understand the adaptive threshold for CFAR with our type of data. From Equation \ref{eq:cfar_pfa_XdB}, $P_{fa}$ can be obtained if the number of noise cells $N$ is known as well as the desired decibel level $X_{dB}$ between the power of the CUT and its corresponding noise power. Note that this approach uses the interference power instead of the interference magnitude.\\
The difference in decibel between the power in the target plus interference and the interference in the training set is $X_dB=0.75dB$. The target+interference (whistler) is 0.2s long, however, only the peak of these cells are of interest. Also, since an adaptive threshold detector is used, $X_dB$ can be decreased in an attempt to increase the probability of detection while maintaining a low probability of false alarm. Therefore, a value of $X_db=0.5dB$ is chosen as for the CFAR threshold. This corresponds to a threshold of $T=29.81dB$ and is associated with a probability of false alarm $P_{fa}=3.36\mathrm{E}{-1}$ and a probability of detection of $P_D=6.82\mathrm{E}{-1}$ as shown in Figure \ref{fig:ch5_detector_stats}. \\

\begin{figure}[!h]
	\centering
	\begin{subfigure}[c]{.8\textwidth}
		\centering
		\includegraphics[width=\textwidth]{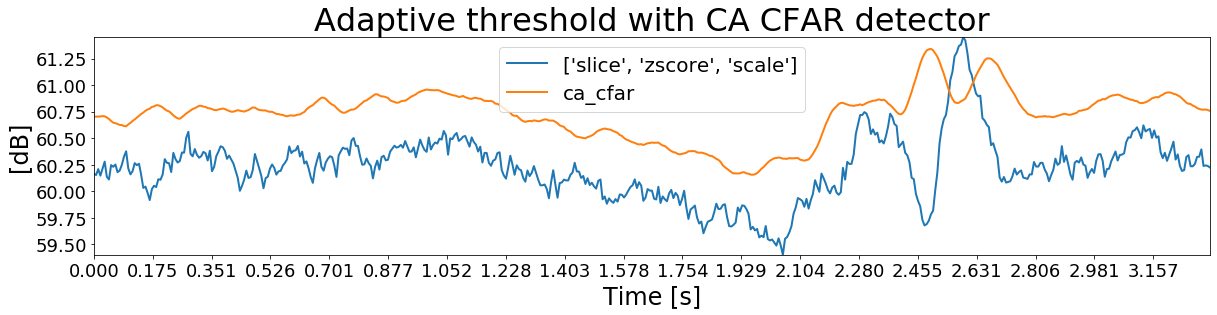}
		\captionsetup{justification=centering}
		\caption{CA CFAR threshold level.}
		\label{fig:ch5_cfar_cacfar_corr}
	\end{subfigure}
	\begin{subfigure}[c]{.8\textwidth}
		\centering
		\includegraphics[width=\textwidth]{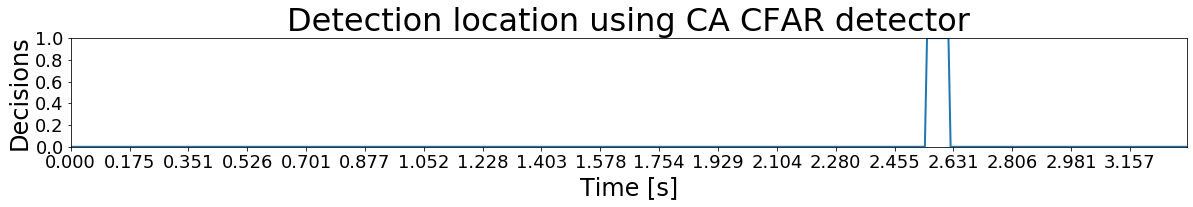}
		\captionsetup{justification=centering}
		\caption{CA CFAR detection decisions.}
		\label{fig:ch5_cfar_cacfar_decisions}
	\end{subfigure}
	\caption{Results of CA CFAR with $N=10$, $G=7$, and $P_{fa}=3.36\mathrm{E}{-3}$. One whistler is detected at $t=2.59s$.}
	\label{fig:ch5_cfar_cacfar}
\end{figure}

The proposed adaptive detector combines three CFAR detectors.\\
The first CFAR detector is the Cell Averaging CFAR (CA CFAR) which uses the noise signal in the noise cell and $X_{dB}$ to set the threshold for the CUT. An example of this CFAR on a sample from the training set with $N=10$, $G=7$, and $P_{fa}=3.36\mathrm{E}{-3}$ (for $X_{dB}=8dB$) is shown in Figure \ref{fig:ch5_cfar_cacfar}.
The CA CFAR detector was able to detect only one out of the three whistlers (t=2.37, t=2.38, t=2.59) present in the spectrogram. The first two whistlers are not detected as they interfere with one another. The detection of these whistlers might be achieved by reducing the window size such that there are no interfering targets, however, this might also increase the probability of false alarm since the noise is not homogeneous.
\\
Order Statistic CFAR (OS CFAR) select the noise power in the $k^{th}$ ordered noise cell as average noise power to be considered by the CFAR detector. Let $\mathcal{N}=\{n_1^2, n_2^2, \ldots, n_N^2, \ldots, n_{2N}^2\}$ be the set of ordered noise power of the noise cells, $x_{CUT}^2$ the power of the CUT, and $\mathcal{\hat{N}}$ the average noise power. Assuming that the noise around the targets have a Gaussian distribution with $x_{target}^2 \geq \mathcal{\hat{N}}+X_{dB}$, since $\mathcal{\hat{N}}$ is ordered, $n_k^2$ for $k=N$ is likely to be equal to $\mathcal{\hat{N}}$. If either side of the CUT has at least one interfering target \footnote{Interfering target, in this case, refers a cell belonging to another whistler.}, it is safer to assume that the interfering target[s] covers the half side of the noise cells. Consequently, $k \leq N$ is a preferred option. If the CUT is surrounded by interfering targets on both sides, changing the value of $k$ will not help, instead, a different window size should be chosen to ensure that such case do not occur.\\
The last CFAR detector is the Trimmed Mean CFAR (TM CFAR). Trimmed Mean CFAR (TM CFAR) can be though of as a generalisation of OS CFAR. Instead of a single noise cell used to calculate the noise power, a section of the ordered noise power in each noise cell is used. TM CFAR requires two additional parameters $T_S$ and $T_L$ on top of $N$, $G$ and $P_{fa}$. $T_S$ and $T_L$ are used to select the subset $\mathcal{\hat{N}}_{(L,S)}$ of $\mathcal{\hat{N}}$ such that $\mathcal{\hat{N}}_{(L,S)} = \{n_{T_S}^2, \ldots, n_{T_L}^2\}$ with $0 \leq T_S < T_L \leq 2N$. $T_L$ is chosen assuming that the CUT is surrounded by interfering targets, while $T_S$ is chosen to reduce clutter edge false alarm.\\

\begin{figure}[ht] 
	\centering    
	\includegraphics[width=0.7\textwidth]{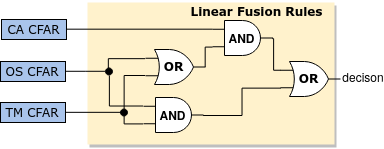}
	\captionsetup{justification=centering}
	\caption{Block Diagram of LF-CFAR.}
	\label{fig:ch2_lf_cfar}
\end{figure}

These detector are combined to form a Linear Fusion CFAR (LF CFAR). An LF CFAR detector as described by Ivkovic \textit{et. al}\cite{Ivkovic2016}  combines the decisions of three CFAR detector (CA CFAR, OS CFAR, and TM CFAR) to decide on the presence of targets in the CUT. LF CFAR makes a new decision by using the logic described in Figure \ref{fig:ch2_lf_cfar}. 
When CA CFAR is 0, there is a possibility that this is due to multiple interfering targets or strong clutter. Thus an AND logic is applied between OS CFAR and TM CFAR.  When CA CFAR is 1, there is a possibility of false alarm caused by either a change in the clutter features or multiple targets interference. Consequently, an AND logic is applied on the CA CFAR output and the result of the OR logic between the output of OS CFAR and TM CFAR since both are designed considering multiple targets interference.

\subsection{Output}
\begin{figure}[!h] 
	\centering
	\includegraphics[width=.8\textwidth]{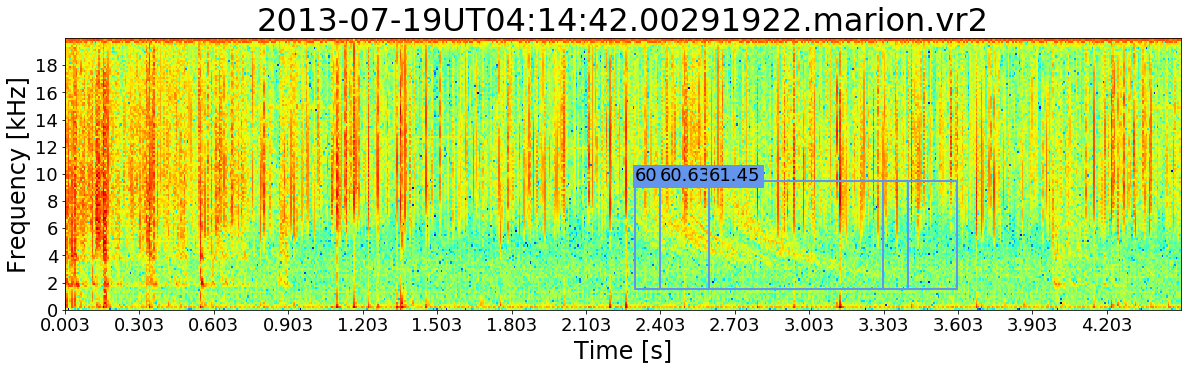}
	\captionsetup{justification=centering}
	\caption{Result of detection using Cross-correlation with a simulated whistler. Three whistlers are detected starting at t=2.3s, t=2.4s, and t=2.6, all with a duration of 1s.}
	\label{fig:ch5_detector_output}
\end{figure}

The output of the detector is the starting time and its associated cross-correlation value for each detected whistler in the spectrogram. The maximum number of whistlers that can be detected in an interval of time is given by the time resolution of the spectrogram; for Marion, the time resolution is $6.4ms$. This leads to results that are sometimes unreadable, thus a time resolution of $0.1s$ is selected. Therefore, whistlers detected within $0.1s$ interval are considered as one single detection with the cross-correlation value being the highest of them all. To satisfy requirement $R_{12}$ from Table \ref{tab:ch3_system_req1}, the ending, therefore, the duration of the whistlers should also be found. This requirement can easily be met by using the duration (1s) of the simulated whistler use as kernel as the duration of each detected whistler. However, a more suitable approach can be used by and taking the duration of the simulated whistler associated with the maximum cross-correlation between a 1s cut of the detected whistler and simulated whistlers of all possible duration.\\
The final result of the detection using the method described in this chapter is shown in Figure \ref{fig:ch5_detector_output}.

\section{Detection Using A Sliding Deep Neural Convolutional Neural Network}
\begin{figure}[!h] 
	\centering
	\includegraphics[width=0.8\textwidth]{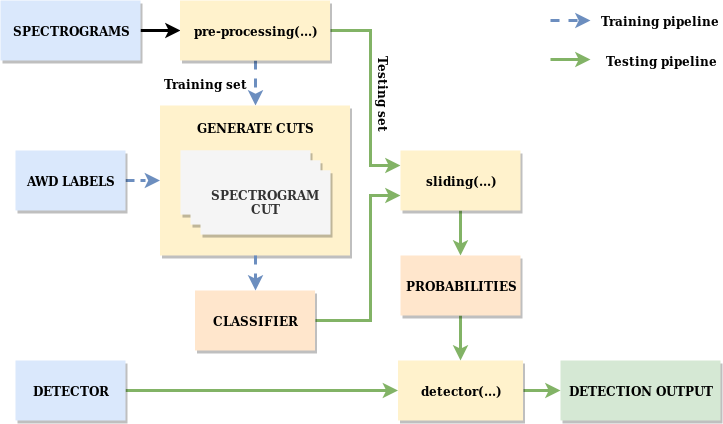}
	\captionsetup{justification=centering}
	\caption{Overview of the whistler detector design using a neural network classifier.}
	\label{fig:ch6_class_design}
\end{figure}

In the domain of computer vision, image classification is the action of identifying objects, people, actions, \ldots from an image. In this section, the design of a whistler detector that uses a whistler classifier to detect and localise whistlers in a spectrogram is presented.\\
The overview of the design of the whistler wave detector using a neural network classifier is presented in Figure \ref{fig:ch6_class_design}. 

The whistler wave detector has two pipelines, a training, and a testing pipeline. During training, a neural network classifier is trained using preprocessed samples from the training set and their AWD labels. During testing/evaluation, the trained classifier is slid over a preprocessed spectrogram resulting in a probability of detection over time curve. This curve is then fed to a detector which detects and locates the whistlers.\\

The first part of designing this detector is training the classifier. This neural network classifier requires sufficient labelled samples for training and a choice of network architecture. The design decisions concerning these two requirements are thoroughly elaborated in this section. 

\subsection{Dataset}
From the Marion dataset, 1471 samples were selected to be from the training set. Using the AWD method, we generate whistler and noise cuts of 1s between 1.5kHz, and 9.5kHz. The noise cuts are the portion of the spectrogram which do not contains whistlers. From this training set, a total of 11148 cuts are generated with 8317 (74.61\%) whistler cuts and 2831 (25.39\%) noise cuts. 

\subsection{Neural Network Architecture}
\begin{figure}[ht] 
	\centering    
	\includegraphics[width=\textwidth]{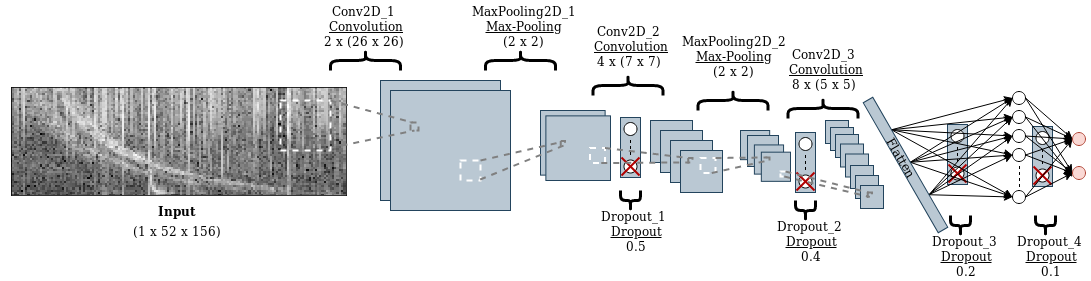}
	\captionsetup{justification=centering}
	\caption{Final architecture of the deep convolutional neural network classifier with additional Dropout layers.}
	\label{fig:ch6_class_cnn_final}
\end{figure}

Numerous neural network architectures have been developed since the popularisation of artificial intelligence \cite{tch_2017}. Among those, Deep Convolutional Neural Networks are known to be adequate for image classification mostly because of their convolutional layers. These networks have two main components, a convolutional base and a classifier. The convolutional base typically features convolution cells or pooling layers and few kernels. This base learns hierarchical feature representations of the data. The object of interest for this project are whistlers. As explained in the literature, whistlers have a simple shape, consequently, a convolutional neural network with at least three convolutional layers should be sufficient for whistlers wave classification. \\
The architecture for the classifier is presented in Figure \ref{fig:ch6_class_cnn_final}.
This network takes each cut as input images with shape $(1 \times 52 \times 156)$ and an output with shape $(2 \times 1)$. The convolutional base is composed of two pairs of convolutional layer - max-pooling layer followed by a convolutional layer. The output of the convolutional base is flattened is fed to the classifier with has one dense layer of 128 neurons and a final dense layer with only two neurons as the output of the classifier using a \textit{softmax} activation. To prevent overfitting, four Dropout layers are added to this network. The entire network has a total of 522,112 trainable parameters.

\subsection{Classifier performance}
\begin{figure}[ht] 
	\centering    
	\includegraphics[width=0.8\textwidth]{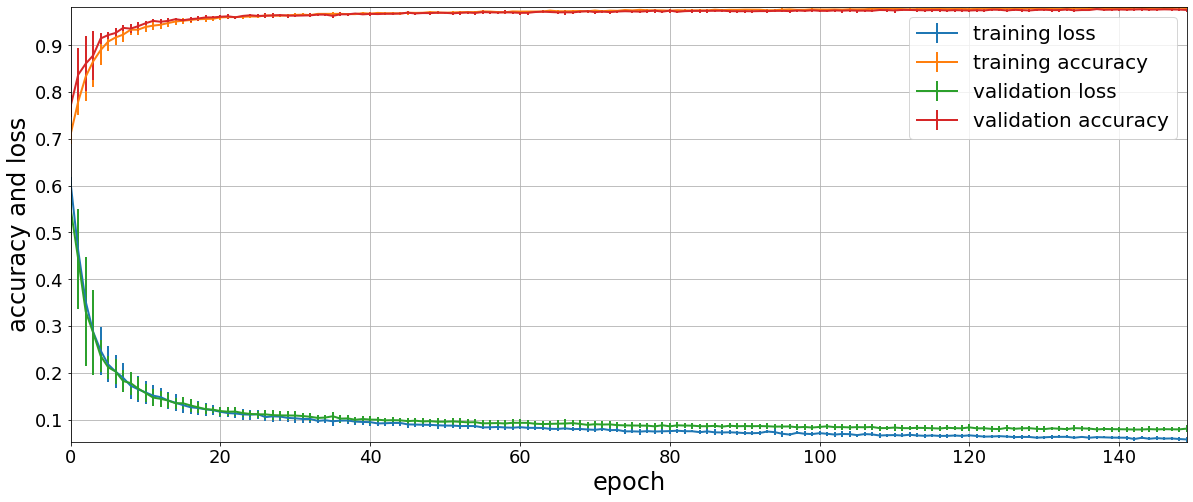}
	\captionsetup{justification=centering}
	\caption{Accuracy and loss of the training using 5 fold stratified shuffle split cross-validation on 7439 training samples and 3679 validation samples.}
	\label{fig:ch6_class_results}
\end{figure}

The model in Figure \ref{fig:ch6_class_cnn_final} is trained in batches of 250 samples from a dataset of 11148 samples comprising 8317 whistler cuts and 2831 noise cuts. The dataset was divided into 5 folds of 150 epochs each with a validation set percentage of 33\% (7469 samples for training, 3679 samples for validation) following the stratified shuffle split cross-validation techniques. We use a Stochastic Gradient Descent (SGD) with a learning rate of 0.001 and a momentum of 0.9 as the optimiser. The loss and accuracy of the training are shown in Figure \ref{fig:ch6_class_results}.
The plots in Figure \ref{fig:ch6_class_results} are the curves with error bars of the accuracy and loss of both the training and validation set of the model. The error in the curves decreases as training continues meaning that the accuracy and loss of both training and validation set converge to the same constant value at each fold. Moreover, both training and validation accuracy converge to 1 after the $80^{th}$ epoch, while the training and validation loss do not converge to the same value but diverge after the $50^{th}$ epoch. To reduce the effect of this divergence, the model should be trained for $100^{th}$.

\begin{figure}[!h]
	\centering
	\begin{subfigure}[t]{.32\textwidth}
		\centering
		\includegraphics[width=\textwidth]{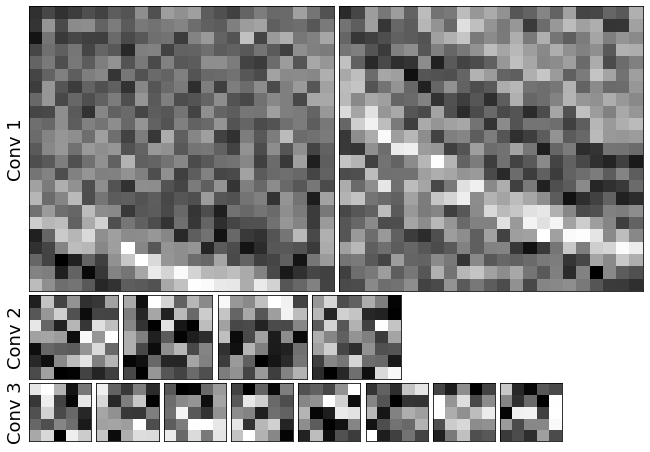}
		\captionsetup{justification=centering}
		\caption{Filters of the trained neural network}
		\label{fig:ch6_class_filters}
	\end{subfigure}
	\begin{subfigure}[t]{.32\textwidth}
		\includegraphics[width=\textwidth]{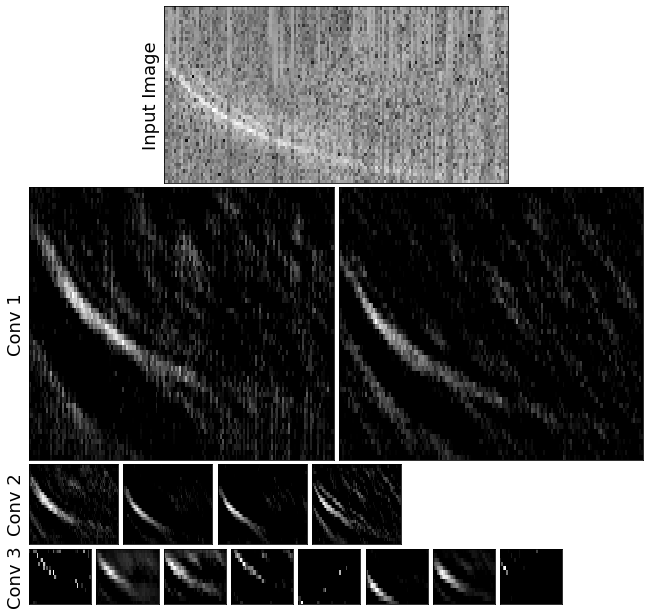}
		\captionsetup{justification=centering}
		\caption{Feature maps when detecting a whistler cut}
		\label{fig:ch6_class_feature_maps_whistler}
	\end{subfigure}
	\begin{subfigure}[t]{.32\textwidth}
		\includegraphics[width=\textwidth]{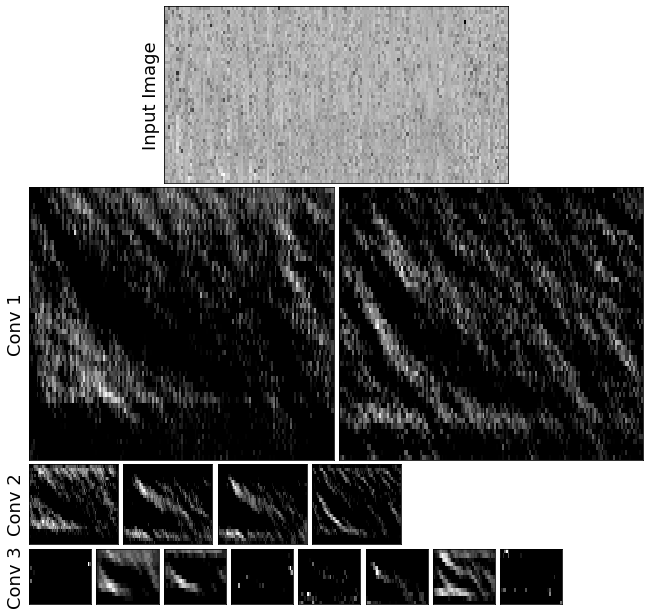}
		\captionsetup{justification=centering}
		\caption{Feature maps when detecting noise cut}
		\label{fig:ch6_class_feature_maps_noise}
	\end{subfigure}
	\captionsetup{justification=centering}
	\caption{Filters and Feature Map of the trained model.}
	\label{fig:ch6_class_kernels}
\end{figure}

The convolutional layers provide insight into the decisions taken by the network. Figure \ref{fig:ch6_class_kernels} shows the convolutional filters learnt by the model as well as their associated feature maps for a whistler cut and a noise cut. 
The two filters of the first convolutional layer (Figure \ref{fig:ch6_class_filters}) show the most basic shapes that are looked for by the model in an image. These filters both have the same feature (line with decreasing gradient) but at a different location in the filters. The similarities in the features can be because only whistlers are present in the whistler cuts and no distinct shapes are present in the noise cuts. As we go deeper into the network, the filters have a more complex shape that is not easily understood because of the low resolution of these filters. The feature maps, however, present more information on the model's prediction. Figure \ref{fig:ch6_class_feature_maps_whistler} and \ref{fig:ch6_class_feature_maps_noise} show the feature maps of the model when it is respectively presented with a whistler cut and noise cut. As expected, both feature maps of the first convolutional layer present the feature corresponding to lines of decreasing gradient in the image. In Figure \ref{fig:ch6_class_feature_maps_whistler}, the shape of the original whistler can be clearly observed while in Figure \ref{fig:ch6_class_feature_maps_noise}, there are no distinct shapes being observed. These features are more and more refined into different new features as we move to the deeper layers of the network.  

\subsection{Sliding classification}
A challenge of object detection is to locate an unknown number of object in an image. This problem cannot be solved directly as an object classification problem since a classifier has a fixed input shape. Let assume that our spectrogram after preprocessing is a $M\times P$ image, the classifier size was initially chosen such that it classifies images of size $M\times N$ with $N<P$. The spectrogram image cannot then be classified directly, instead, we generate a region of interests (RoI) with sizes that match the classifier. Since both spectrogram and classifier have the same width, the RoI are extracted by sliding the classifier along the length of the spectrogram. The resulting number of RoI $N_{RoI}$ is then given by:
\begin{equation}
N_{RoI} = \lfloor \frac{P-N}{S}\rfloor
\end{equation}
with the stride $S$, the distance between two consecutive RoI. The finest resolution can be obtained with $S=1$ when $N_{RoI} = P-N$ and the ROI are distant by the spectrogram time resolution. However, the finer the resolution, the more computationally expensive the detection become. To reduce the computation time, $S$ can be chosen such that the ROI are distant by 0.1s. A stride of 1 will be used in the following section to explain the algorithm of the detector. 

\begin{figure}[ht] 
	\centering    
	\includegraphics[width=.8\textwidth]{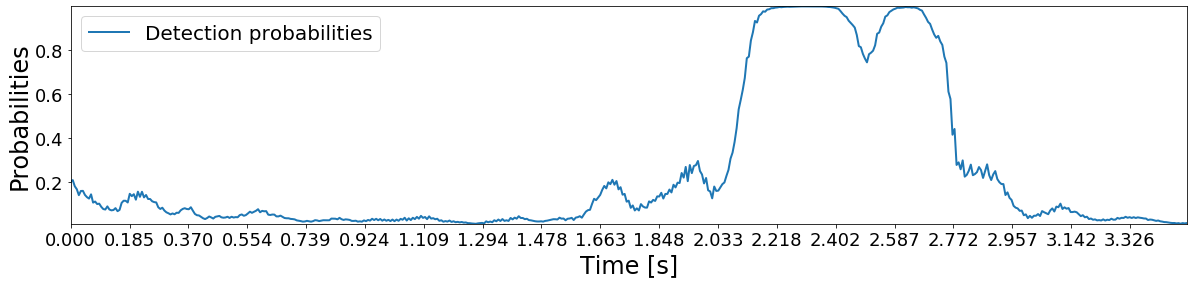}
	\captionsetup{justification=centering}
	\caption{Probabilities of detection of a preprocessed spectrogram}
	\label{fig:ch6_class_spec_probs}
\end{figure}

The result of the classifier on the spectrogram by generating regions of interest with a stride of 1 is shown in Figure \ref{fig:ch6_class_spec_probs}. The result is a curve of the whistler's probability of detection in the preprocessed spectrogram. The probability of detection is below 0.5 in most of the spectrogram except at time between 2 and 2.7 seconds where a series of whistlers are observed). Moreover, Figure \ref{fig:ch6_class_spec_probs} shows the presence of two whistlers with the lowest probabilities at 0.8 and different thickness. 

\subsection{Detector}
\begin{figure}[ht] 
	\centering    
	\includegraphics[width=.8\textwidth]{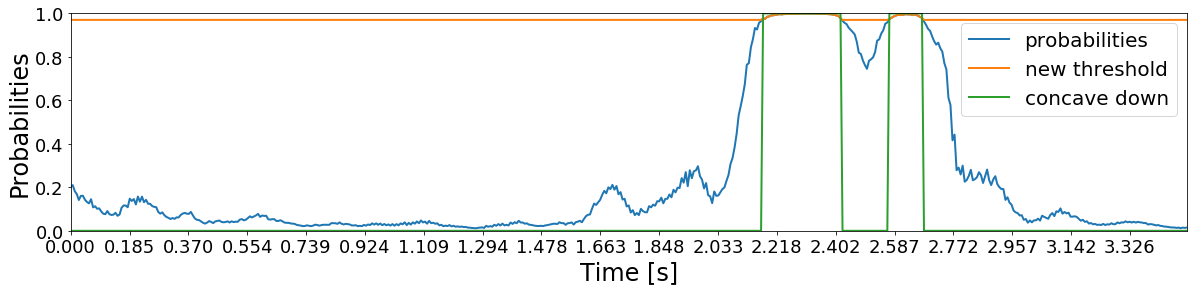}
	\captionsetup{justification=centering}
	\caption{Detector decisions with a new threshold at 0.9 with two whistlers detected at $t=[2.334, 2.622] s$.}
	\label{fig:ch6_class_spec_detections}
\end{figure}

The last phase of the detection is the decision of the whistler location. The classifier performs quite well on the dataset with an accuracy of 0.976 and a loss of 0.08. As a result, finding the peak of a detected whistler is quite hard since a group of whistlers have a detection probability of 1. Consequently, using a static threshold is not the perfect option, instead we find the peak of the whistlers by finding the peaks of the concave down portions of the spectrogram above a probability threshold. For this detector, we select an arbitrary threshold of 0.9. The result of this detection is shown in Figure \ref{fig:ch6_class_spec_detections}. The two whistlers present in the spectrogram are detected at time t=2.334s and t=2.622s. 

\subsection{Detector Output}
\begin{figure}[ht] 
	\centering    
	\includegraphics[width=.8\textwidth]{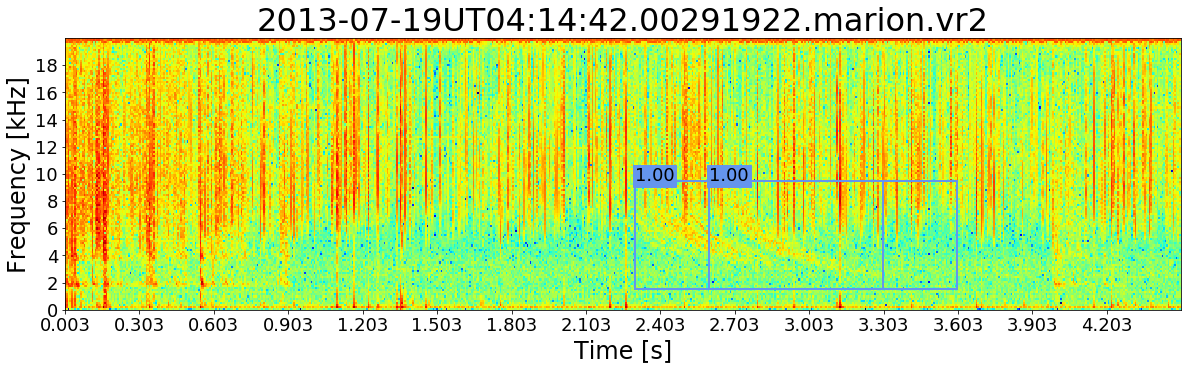}
	\captionsetup{justification=centering}
	\caption{Bounding box prediction using the sliding deep convolutional neural network classifier.}
	\label{fig:ch6_class_output}
\end{figure}

After finding the starting location of whistlers, the ending location is found using the length of the whistler cut use for training the data. The final result of the detection for a sample from the training set showing the bounding box prediction with the probability of detection is shown in Figure \ref{fig:ch6_class_output}.
Two whistlers are found at t=2.3 and t=2.6 seconds. All detected whistlers have similar bounding boxes since the frequency range and the end time are static (1.5kHz to 9.5kHz and a duration of one second).

\section{Detection Using You Only Look Once}
\begin{figure}[ht] 
	\centering    
	\includegraphics[width=.8\textwidth]{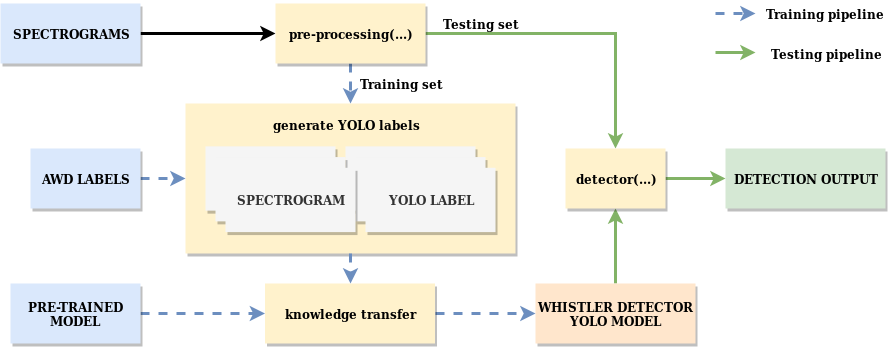}
	\captionsetup{justification=centering}
	\caption{Overview of the whistler detector design using a YOLO model.}
	\label{fig:ch6_yolo_design}
\end{figure}

You Only Look Once is the current state of the art in object detection. In this section, we adapt the implementation of YOLO to create a whistler detector and localiser.\\
The design proposed in this chapter make uses of the state of the art in object detection as a method of detecting whistlers in a spectrogram. An overview of the design is presented in Figure \ref{fig:ch6_yolo_design}. This design has two pipelines, a training stage,  and testing pipeline. The training pipeline has three parts during which preprocessed samples from the training set are used to generate YOLO formatted labelled using the AWD labels and are later used to train a pre-trained YOLO model. The testing pipeline has two parts in which the preprocessed samples from the testing set are evaluated on the new YOLO model.

In the following sections, the choice of the inputs to the pipeline as well as the design decision for the processes are elaborated.

\subsection{YOLO dataset generation}
YOLO is a supervised machine learning algorithm which uses bounding boxes and class label as label data. YOLO requires five information from every labelled object in the form of $(c,x,y,w,h)$, with $x$ and $y$, the normalised coordinate of the centre of the object, $w$ and $h$, the normalised width and height of the object, and $c$ the object class. 

\begin{figure}[ht] 
	\centering    
	\includegraphics[width=.8\textwidth]{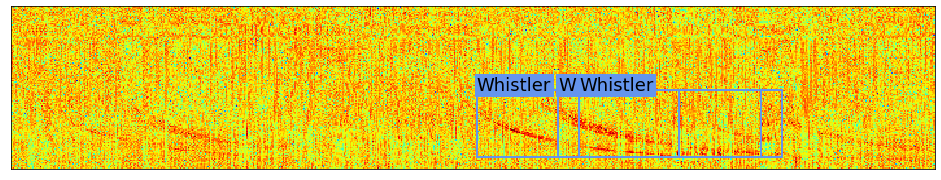}
	\captionsetup{justification=centering}
	\caption{Sample from the labelled dataset used for training the YOLO model. Three whistlers are detected by AWD and the YOLO labels are $[0, 0.58, 0.73, 0.17, 0.45]$, $[0, 0.67, 0.73, 0.17, 0.45]$, and $[0, 0.69, 0.73, 0.17, 0.45]$}
	\label{fig:ch6_yolo_data}
\end{figure}

The labelled dataset generated for training our model using YOLO is obtained from our training set. Each vr2 sample is converted to a spectrogram, preprocessed, converted into an RGB image and then labelled using the AWD by Lichtenberger \cite{Lichtenberger2008}. \\
Figure \ref{fig:ch6_yolo_data} shows a sample from the training set and its YOLO label.

\subsection{Knowledge Transfer}
Knowledge transfer in the domain of machine learning is used to transfer the knowledge acquired by a trained network to a new network\cite{Goodfellow2016} and is based on the assumption that similar tasks require the same underlying knowledge. The common Deep Neural Network (DNN) has two main components, a convolutional base and a classifier. The convolutional base typically learns hierarchical feature representation of the data, - according to Yosinski\cite{Yosinski2014}, in such cases, there is a general to specific learning process occurring in the network, - while the classifier is generally a set of fully connected layers that classify an image according to its features.
The first layers of the network thus contain features that are domain-independent while the last layers contain domain-dependent features. There are two main strategies in knowledge transfer known as fine-tuning and transfer learning. Fine-tuning consists of continuing training a pre-trained network. This strategy is computationally expensive especially for very deep neural networks. The second strategy is to freeze the first few layers of the pre-trained network before training. This results in faster training time at the expense of the network performance. To reduce training time during experimentation, we use transfer learning to train the pre-trained YOLO model.

\subsection{YOLO Model selection}
Few pre-trained YOLO models are available and among those trained by Redmon \textit{et al.} \cite{yolov3}, YOLOv3-spp and YOLOv3-tiny are the one of interest. The performance of these model is presented in Table \ref{tab:ch6_yolo_weights}
\begin{table}[!h]
	\centering
	\caption{Performance on the COCO Dataset}
	\label{tab:ch6_yolo_weights}
	\begin{tabular}{@{}lrrrrrr@{}}
		\toprule
		\textbf{Model} & \textbf{Train} & \textbf{Test} & \textbf{mAP} & \textbf{FLOPS} & \textbf{FPS} & \textbf{Size} \\ 
		\midrule
		YOLOv3-tiny & COCO trainval & test-dev & 33.1 & 5.56 Bn & 220 & 34M \\
		YOLOv3-spp & COCO trainval & test-dev & 60.6 & 141.45 Bn & 20 & 241M\\
		\bottomrule
	\end{tabular}
\end{table}

YOLOv3-spp is the latest YOLO pre-trained model and is more accurate (60.6 mAP) than YOLOv3-tiny but at the expense of more computations (26 times more than YOLOv3-tiny). YOLOv3-tiny will be the model of choice during experimentation. 

\subsection{Detector Output}
\begin{figure}[!h] 
	\centering
	\includegraphics[width=.8\textwidth]{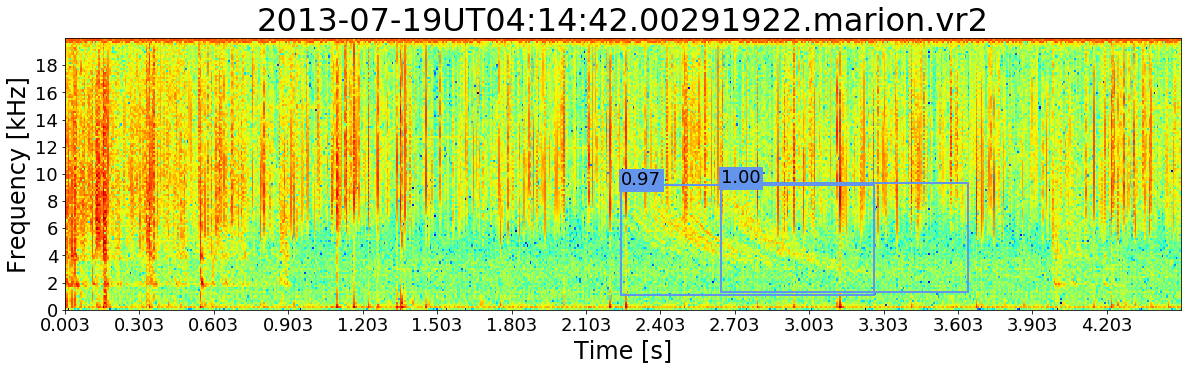}
	\captionsetup{justification=centering}
	\caption{Result of detection using YOLO with a simulated whistler. Two whistlers are detected at t=2.24s and t=2.64s.}
	\label{fig:ch6_yolo_output}
\end{figure}

Once the whistler detector is trained, the result of the detection is a list of bounding boxes in YOLO format along with the class label and class probabilities. The result of a detector trained using the training set with non-processed sample for the sample in Figure \ref{fig:ch5_detector_output} is shown in Figure \ref{fig:ch6_yolo_output}. The YOLO network detected two whistlers with a probability of 0.97 and 1, starting time of 2.24 and 2.64 seconds and duration of 1s. The object of interest for this project are whistlers. As explained in the literature, whistlers have a simple shape, consequently, a convolutional neural network with at least three convolutional layers should be enough to classify the whistlers from the noise. 

\section{Model performance and validation}
The Automatic Whistler Detector developed by Lichtenberger \textit{et al.} \cite{Lichtenberger2008} has been used to generate the labels of the data used for this research. His method outputs the 5kHz time of the detected whistler. In this research, the starting and ending time of each whistler are of interest. Therefore, we extract the starting point of the whistler located at 5kHz using the statistic obtained from whistlers in the training set. Using this starting point can be misleading in the evaluation of the performance. Instead, the detector is evaluated based on the number of detected series of whistler. Therefore, the following metrics are defined:
\begin{itemize}
	\item \textbf{True Positive $TP$}: The number of whistler found by AWD that are within the bounding boxes of our detection.
	\item \textbf{False Positive $FP$}: The number of bounding boxes which do not contain any whistlers detected by AWD. 
	\item \textbf{False Negative $FN$}: The number of whistlers found by the AWD that are not within any of our bounding boxes.
	\item \textbf{True Negeative $TN$}: The noise detected by both AWD and the new methods. This is not applicable for this research since we are not interested in detected the noise.
\end{itemize}

\subsection{Best Model Performance}
The design and implementation of a few whistler detector and localiser was presented. These designs, however, are not the ones achieving the best performance of both the training set of the data collected at Marion. In this section, the parameters of each design is altered to obtain the best performance on the training data.
\subsubsection{Cross-Correlation with a Simulated Whistler}
This CCSW design has three inputs whose parameters can be altered to find the best model. For the spectrogram, the 1D preprocessing method can be changed. The kernel's frequency range $f$ and zero-dispersion $D_0$ can be adjusted. And finally, the parameters of the detector ($N$,$G$,$X_{dB},$$k$,$T_S$,$T_L$) can be adjusted.
Table \ref{tab:ch7_CCSW_marion_performance} presents the performance of the best models on Marion's training set.
\begin{table}[!h]
	\centering
	\caption{Performance of CCSW detector with different tunning parameters.}
	\label{tab:ch7_CCSW_marion_performance}
	\begin{tabular}{@{}cccccc@{}}
		\toprule	
		\multicolumn{3}{c}{\textbf{Parameters}} & \multicolumn{3}{c}{\textbf{Performance}} \\ 
		\midrule
		Preprocessing & \begin{tabular}[c]{@{}c@{}}Kernel \\ {[}$f_{min},f_{max}${]}\end{tabular} & \begin{tabular}[c]{@{}c@{}}Detector\\ (N,G,$X_{dB}$,k,$T_S$,$T_L$)\end{tabular} & Misdetection & False Alarm & \multicolumn{1}{l}{F1 score} \\
		\midrule
		None & {[}1.5,9.5{]} & (10,5,0.3,13,5,7) & 0.092 & 0.199 & 0.851 \\
		Zscore & {[}1.5,9.5{]} & (12,10,0.5,13,5,3) & 0.106 & 0.108 & 0.893 \\
		Zscore & {[}4.5,11.5{]} & (15,7,0.7,15,2,7) & 0.045 & 0.174 & 0.886 \\
		C. Detrend & {[}1.5,9.5{]} & (8,5,0.4,5,0,7) & 0.062 & 0.159 & 0.887 \\
		L. Detrend & {[}1.5,9.5{]} & (12,10,0.4,13,8,5) & 0.088 & 0.124 & 0.894\\
		\bottomrule
	\end{tabular}
\end{table}
All detectors achieve a good f1 score of 0.882 on average on Marion's training set. Without any preprocessing, the detector achieved the worst overall performance with an f1 score of 0.851. A kernel with the frequency range between 4.5 and 11.5 kHz as selected in the AWD results in a good misdetection rate but a high false alarm. Preprocessing using either the Z score, a constant or linear detrend with $1.5 \leq f \leq 9.5$ results in similar performances on both training and testing set. However, the Z score performs quite well with both misdetection and false alarm of about 10\% for the training set.\\
\subsubsection{Sliding Deep Convolutional Neural Network (SDCNN)}
The SDCNN requires the generation of regions of interest and a trained classifier. In this section, the performance of this detector is evaluated for different preprocessing techniques with RoI generated such that two consecutive RoI from a spectrogram is distant by 0.1s. For each preprocessing technique, the classifier in Figure \ref{fig:ch6_class_cnn_final} was trained for 100 epochs. The results of the evaluation on Marion's training set is tabulated in Table \ref{tab:ch7_sdcnn_performance}. 
\begin{table}[!h]
	\centering
	\caption{Performance of the sliding deep convolutional neural network on Marion's training set.}
	\label{tab:ch7_sdcnn_performance}
	\begin{tabular}{@{}cccccc@{}}
		\toprule
		\textbf{Parameters} &
		\multicolumn{2}{c}{\textbf{Classifier}} & \multicolumn{3}{c}{\textbf{Performance}} \\ \midrule
		Preproessing & Accuracy & Loss & Misdetection & False Alarm & F1 score \\ \midrule
		None & 0.974 & 0.087 & 0.089 & 0.04 & 0.935 \\
		Zscore & 0.977 & 0.077 & 0.092 & 0.036 & 0.935 \\
		C.Detrend & 0.979 & 0.076 & 0.102 &  0.03 & 0.932 \\
		L. Detrend & 0.97 & 0.096 & 0.109 & 0.025 & 0.931 \\
		\bottomrule
	\end{tabular}
\end{table}

Table \ref{tab:ch7_sdcnn_performance} shows the accuracy and loss of the classifier during training as well as the performance of the detector on Marion's training set. For each preprocessing technique, the classifiers reached an accuracy and a loss of respectively about 0.975 and 0.084 on average. All detectors achieved a high f1 score around 0.933 with a misdetection rate between 0.089 and 0.109 and a false alarm rate below 0.04 on the training set.\\
The best detectors are obtained when spectrograms are not preprocessed or when preprocessed with the Z score transform. Both detectors have the same f1 score but the Z score has slightly higher misdetection and a lower false alarm rate.\\
The evaluation of the detectors on Marion's testing set produces a similar result to the training set. Once again, the Z score performs quite well with an overall 0.939 for the f1 score with one of the lowest false alarm and misdetection rate.
\subsubsection{You Only Look Once (YOLO)}
After training the YOLO model on Z score preprocessed spectrograms, the same model is trained on data preprocessed with different other techniques. The results of the evaluation of these new detectors on Marion's training set are tabulated in Table \ref{tab:ch7_yolo_performance}. 
\begin{table}[!h]
	\centering
	\caption{Performance of the sliding deep convolutional neural network on Marion's training set.}
	\label{tab:ch7_yolo_performance}
	\begin{tabular}{@{}ccccc@{}}
		\toprule
		\textbf{Parameters} & \textbf{YOLO} & \multicolumn{3}{c}{\textbf{Performance}} \\ 
		\midrule
		Preprocessing & Loss & Misdetection & False Alarm & F1 score \\ 
		\midrule
		None & 0.351 & 0.085 & 0.002 & 0.955  \\
		Zscore & 0.352 & 0.075 & 0.001 & 0.961  \\
		C. Detrend & 0.361 & 0.109 & 0.002 & 0.942  \\
		L. Detrend & 0.364 & 0.057 & 0.001 & 0.97  \\
		\bottomrule
	\end{tabular}
\end{table}

After 280000 iterations, the models' loss was 0.357 on average with the linear detrend techniques having the highest loss. All detectors perform very well on Marion's training set with an average of 0.082, 0.002, and 0.957 respectively for the misdetection, false alarm, and f1 score. Among these detectors, the ones with the Z score and the linear detrend have the highest performances, however, the latter achieves a lower misdetection and false alarm rate on the training set. The model trained using preprocessed samples with linear detrend is, therefore, the best model.

\subsection{Models comparison}
All detectors each outputs the results of the whistlers' detection and localisation given an vr2 file with each detection having a figure of merit. These detectors thus meet requirement $R_1$ (Table \ref{tab:ch3_system_req1}).\\ 
The best detectors from the three designs developed in this study are evaluated on Marion's testing set.  The performance of these detectors are tabulated in   Table \ref{tab:ch7_model_comparison_marion}.
\begin{table}[!h]
	\centering
	\caption{Performance of each detector on Marion's testing set.}
	\label{tab:ch7_model_comparison_marion}
	\begin{tabular}{@{}cccc@{}}
		\toprule
		\multirow{2}{*}{\textbf{Detector}} & \multicolumn{3}{c}{\textbf{Performance}} \\ 
		& Misdetection & False Alarm & F1 score \\ 
		\midrule
		CCSW  & 0.096 & 0.094 & 0.906  \\
		SDCNN  & 0.093 & 0.026 & 0.939  \\
		YOLO  & 0.141 & 0.021 & 0.915  \\ 
		\bottomrule
	\end{tabular}
\end{table}
All detectors perform very well on Marion's testing set with the poorest performance with an f1 score of 0.096 from the CCSW. The best performance is achieved by the SDCNN with an f1 score of 0.939, a misdetection rate of 0.093 and a false alarm rate of 0.026. All detector satisfy specification $S_2$ from Table \ref{tab:ch3_system_specs} with all misdetection and false alarm below 20\%. \\
The processing time of the detectors was also evaluated by computing the processing ratio $r$ (see requirement 2 in Table \ref{tab:ch3_system_req2}), defined as the ratio between the processing time of a detector on a sample and the duration of this sample. This evaluation was performed on 5 batches of 50 samples from Marion's training set. The results of the evaluation is tabulated in Table \ref{tab:ch7_model_speed}. 
\begin{table}[!h]
	\centering
	\caption{Ratio of processing time over sample duration for each detector.}
	\label{tab:ch7_model_speed}
	\begin{tabular}{@{}rccc@{}}
		\toprule
		& \textbf{CCSW} & \textbf{SDCNN} & \textbf{YOLO} \\ 
		\midrule
		$r$  & 0.179 & 0.376 & 0.010 \\
		\bottomrule
	\end{tabular}
\end{table}
All detectors have a ratio below 0.5, this indicates each detector takes less than half of the data collection time to process the data. This satisfies requirement $R_2$ and specification $S_3$.\\

The detectors were also evaluated on SANAE IV's dataset despite the poor quality of this dataset. Table \ref{tab:ch7_model_comparison_sanae} presents the performance of the detector on SANAE IV's dataset.
\begin{table}[!h]
	\centering
	\caption{Performance of each detector on SANAE IV's dataset.}
	\label{tab:ch7_model_comparison_sanae}
	\begin{tabular}{@{}cccc@{}}
		\toprule
		\multirow{2}{*}{\textbf{Detector}} & \multicolumn{3}{c}{\textbf{Performance}} \\ 
		& Misdetection & False Alarm & F1 score \\ 
		\midrule
		CCSW  & 0.91 & 0.157 & 0.04  \\
		SDCNN  & 0.532 & 0.055 & 0.626  \\
		YOLO  & 0.815 & 0.043 & 0.31  \\
		\bottomrule
	\end{tabular}
\end{table}
The performance of the detectors on SANAE IV's dataset is worst compared to the one on Marion's testing set. The false alarm rate is still below 20\%, so more than 80\% of the detections were correct. However, the misdetections are considerably high with the worst performance reaching a misdetection of 0.815 and 0.91. Yet, the best performance is achieved by the SDCNN with a misdetection of 0.532, a false alarm rate of 0.055 and an f1 score of 0.626. \\
The corruption of SANAE IV's dataset participated in degrading the detector's performance, however, the whistlers present in this data differ from the ones in Marion. 
As opposed to the ones present in Marion's dataset which correspond to a simulated whistler of zero-dispersion $D_0$ of 80, SANAE IV's whistlers correspond to simulated whistlers with a zero-dispersion of approximately 35 using the Bernard approximation. In fact simply by solely changing $D_0$ in CCSW from 80 to 35, the detector achieves an f1 score of 0.597 which is an improvement of 0.593.

\section{Conclusion and Recommendations}

The goal of this research was to design and implement new methods to detect and localise whistler radio waves that would meet the project requirements given in Tables \ref{tab:ch3_system_req1} and \ref{tab:ch3_system_req2}. This research results in the design and implementation of three different detectors. \\
The first detector (CCSW) gets its inspiration from the current state of the art in whistler wave detection developed by Lichtenberger \textit{et al.} \cite{Lichtenberger2008}. This method makes use of the whistler's dispersion approximation developed by Bernard \cite{Bernard1973} to simulate a whistler which is cross-correlated with preprocessed spectrograms. The results is fed to an adaptive threshold detector to detect the starting point of the whistler.\\
The second detector (SDCNN) uses a deep convolutional neural network as a classifier to detect the presence of whistler radio waves on generated Region of Interest (RoI).\\
The last detector (YOLO) uses the state of the art in object detection developed by Redmon \textit{et al.} \cite{Redmon2016}. This detector is an adaptation of YOLO on the data available for this research. The main challenge with this detector is its failure to distinguish very close objects. Moreover, it requires the training data to be labelled, with the labels being the bounding boxes of the whistlers in each spectrogram image. However, the labels provided by AWD do not provide such information, and therefore, an approximation was made. This approximation which generalised the bounding boxes introduced errors in the label which also affected the performance of this detector.\\

The recommendations below are made from the design, the outputs and the performance of the detectors developed in this study.\\
An analysis of the whistlers present in Marion's training set led to the conclusion that most the magnitude of these whistlers are present between 1.5kHz and 9.5kHz of the spectrograms. The cross-correlation kernel used in CCSW uses a simulated whistler over the entire range. Reducing this range to one maximising the magnitude of the whistlers could lead to better performances. Therefore, the effect of the frequency range on the performance of the detectors should be investigated. \\
The characteristics of the whistler extracted from Marion's dataset matches only a certain type of whistlers with zero-dispersion of around 80. However, it was later found that SANAE IV's dataset has whistler with a different characteristic. The lack of diversity in the available data for this research constrained the generalisation of the detectors. It is recommended that for further investigations on whistler waves detection, a diverse dataset containing a few thousand samples from different sites should be available.\\ 
To alleviate problems associated with a lack of diversity in the data or poor labelling, the Bernard approximation of the dispersion of whistler waves could be used to simulate a dataset for the research. Machines learning models such are SDCNN and YOLO could be trained on the diverse simulated dataset and the learnt knowledge transferred on the real dataset training set.\\
The framework used to train the YOLO detector lacked adequate tools for collecting the performance metrics of the training. It is recommended that a complete implementation of YOLO should be done instead of using an Off-the-shelf model. This implementation could abllow fine tunning of the parameters and more control of the detector.

\bibliographystyle{IEEEtran}
\bibliography{main}

\appendix

\section{Detector Performance}
\label{sec:detectors_performance}
\begin{table}[!h]
	\centering
	\caption{Performance of CCWS detector with different tunning parameters.}
	\label{tab:apb_CCWS_marion_performance}
	\begin{tabular}{@{}cccccc@{}}
		\toprule	
		\multicolumn{3}{c}{\textbf{Parameters}} & \multicolumn{3}{c}{\textbf{Performance}} \\ 
		\midrule
		Preprocessing & \begin{tabular}[c]{@{}c@{}}Kernel \\ {[}$f_{min},f_{max}${]}\end{tabular} & \begin{tabular}[c]{@{}c@{}}Detector\\ (N,G,$X_{dB}$,k,$T_S$,$T_L$)\end{tabular} & Misdetection & False Alarm & \multicolumn{1}{l}{F1 score} \\
		\midrule
		\multicolumn{6}{c}{\textbf{Marion Island} (Training Set)} \\
		None & {[}1.5,9.5{]} & (10,5,0.3,13,5,7) & 0.092 & 0.199 & 0.851 \\
		Zscore & {[}1.5,9.5{]} & (12,10,0.5,13,5,3) & 0.106 & 0.108 & 0.893 \\
		Zscore & {[}4.5,11.5{]} & (15,7,0.7,15,2,7) & 0.045 & 0.174 & 0.886 \\
		C. Detrend & {[}1.5,9.5{]} & (8,5,0.4,5,0,7) & 0.062 & 0.159 & 0.887 \\
		L. Detrend & {[}1.5,9.5{]} & (12,10,0.4,13,8,5) & 0.088 & 0.124 & 0.894 \\
		\midrule
		\multicolumn{6}{c}{\textbf{Marion Island} (Testing Set)} \\
		None & {[}1.5,9.5{]} & (10,5,0.3,13,5,7) & 0.096 & 0.199 & 0.851 \\
		Zscore & {[}1.5,9.5{]} & (12,10,0.5,13,5,3) & 0.096 & 0.094 & 0.906 \\
		Zscore & {[}4.5,11.5{]} & (15,7,0.7,15,2,7) & 0.047 & 0.157 & 0.886 \\
		C. Detrend & {[}1.5,9.5{]} & (8,5,0.4,5,0,7) & 0.066 & 0.146 & 0.892 \\
		L. Detrend & {[}1.5,9.5{]} & (12,10,0.4,13,8,5) & 0.086 & 0.116 & 0.899\\ 
		\midrule
		\multicolumn{6}{c}{\textbf{SANAE IV} (Entire dataset)} \\
		None & {[}1.5,9.5{]} & (10,5,0.3,13,5,7) & 0.91 & 0.157 & 0.162 \\
		Zscore & {[}1.5,9.5{]} & (12,10,0.5,13,5,3) & 0.98 & 0.1 & 0.04\\
		C. Detrend & {[}1.5,9.5{]} & (8,5,0.4,5,0,7) & 0.901 & 0.078 & 0.178  \\
		L. Detrend & {[}1.5,9.5{]} & (12,10,0.4,13,8,5) & 0.954 & 0.189 & 0.086\\
		\bottomrule
	\end{tabular}
\end{table}

\begin{table}[!h]
	\centering
	\caption{Performance of the sliding deep convolutional neural network on Marion data.}
	\label{tab:apb_sdcnn_performance}
	\begin{tabular}{@{}cccccc@{}}
		\toprule
		\textbf{Parameters} &
		\multicolumn{2}{c}{\textbf{Classifier}} & \multicolumn{3}{c}{\textbf{Performance}} \\ \midrule
		Preproessing & Accuracy & Loss & Misdetection & False Alarm & F1 score \\ \midrule
		\multicolumn{6}{c}{\textbf{Marion Island} (Training Set)} \\
		None & 0.974 & 0.087 & 0.089 & 0.04 & 0.935 \\
		Zscore & 0.977 & 0.077 & 0.092 & 0.036 & 0.935 \\
		C.Detrend & 0.979 & 0.076 & 0.102 &  0.03 & 0.932 \\
		L. Detrend & 0.97 & 0.096 & 0.109 & 0.025 & 0.931 \\ \midrule
		\multicolumn{6}{c}{\textbf{Marion Island} (Testing Set)} \\
		None & N/A & N/A & 0.102 & 0.036 & 0.93 \\
		Zscore & N/A & N/A & 0.093 & 0.026 & 0.939 \\
		C.Detrend & N/A & N/A & 0.112 & 0.026 & 0.929 \\
		L. Detrend & N/A & N/A & 0.113 & 0.018 & 0.932 \\ 
		\midrule
		\multicolumn{6}{c}{\textbf{SANAE IV} (Entire dataset)} \\
		None & N/A & N/A & 0.568 & 0.07 & 0.59 \\
		Zscore & N/A & N/A & 0.532 & 0.055 & 0.626 \\
		C.Detrend & N/A & N/A & 0.494 & 0.089 & 0.651 \\
		L. Detrend & N/A & N/A & 0.574 & 0.066 & 0.585 \\ 
		\bottomrule
	\end{tabular}
\end{table}

\begin{table}[!h]
	\centering
	\caption{Performance of YOLO on Marion data.}
	\label{tab:apb_yolo_performance}
	\begin{tabular}{@{}ccccc@{}}
		\toprule
		\textbf{Parameters} & \textbf{YOLO} & \multicolumn{3}{c}{\textbf{Performance}} \\ 
		\midrule
		Preprocessing & Loss & Misdetection & False Alarm & F1 score \\ 
		\midrule
		\multicolumn{5}{c}{\textbf{Marion Island} (Training Set)} \\
		None & 0.351 & 0.085 & 0.002 & 0.955  \\
		Zscore & 0.352 & 0.075 & 0.001 & 0.961  \\
		C. Detrend & 0.361 & 0.109 & 0.002 & 0.942  \\
		L. Detrend & 0.364 & 0.057 & 0.001 & 0.97  \\ 
		\midrule
		\multicolumn{5}{c}{\textbf{Marion Island} (Testing Set)} \\
		None & N/A & 0.159 & 0.023 & 0.904  \\
		Zscore & N/A & 0.147 & 0.019 & 0.912  \\
		C. Detrend & N/A & 0.192 & 0.013 & 0.889  \\
		L. Detrend & N/A & 0.141 & 0.021 & 0.915  \\ 
		\midrule
		\multicolumn{5}{c}{\textbf{SANAE IV} (Entire dataset)} \\
		None & N/A & 0.704 & 0.082 & 0.448  \\
		Zscore & N/A & 0.8 & 0.047 & 0.331  \\
		C.Detrend & N/A & 0.766 & 0.061 & 0.375  \\
		L. Detrend & N/A & 0.815 & 0.043 & 0.31  \\ 
		\bottomrule
	\end{tabular}
\end{table}

\newpage
\section*{Acknowledgment}
I was assisted by many people throughout the course of this research.
First and foremost, I would like to express my gratitude to both my supervisors, Dr Amit Mishra from the University of Cape Town (UCT) and Dr Stefan Lotz from the South African Space Agency (SANSA) for their advice and guidance throughout the course of this project. Dr Amit Mishra, for making the time to get weekly updates on the project, for providing direction to the research. Dr Stefan Lotz, for providing the data required for this research and other resources, for introducing me to the Machine Learning Conference in Helio Physic which provided me with a deeper insight into this project.

I would also like to thank Dr Steven Pain and doctor-to-be, Darryn Jordan from the Radar Remote Sensing Group (RRSG) of UCT for their insight on this project and also for providing help with setting up the servers. To the other colleagues from the RRSG, I would like to express my gratitude for the time you have made helping me understand new facets of the research.

I would like to thank the Mandela Rhodes Foundation for their financial and leadership support for the two years of master degree.

Finally, I would like to express my gratitude to friends and family and especially my girlfriend Michaela for her support and for proofreading this dissertation.  
\end{document}